\documentclass[aps, pra, twocolumn, superscriptaddress]{revtex4-2}
\usepackage[utf8]{inputenc}
\usepackage{amsmath}
\usepackage{amssymb}
\usepackage{graphicx}
\usepackage{float}
\usepackage{geometry}
\usepackage{braket}
\usepackage{appendix}
\usepackage{url}
\usepackage{subcaption}
\usepackage{siunitx}
\usepackage{enumitem}
\usepackage{dcolumn}
\usepackage{bm}
\usepackage{ragged2e}
\usepackage{xcolor}
\newgeometry{left=25mm,right=25mm,top=25mm,bottom=20mm}
\setcitestyle{numbers}

\definecolor{myblue}{rgb}{0.12156862745098039, 0.4666666666666667, 0.7058823529411765}
\usepackage{hyperref}
\hypersetup{
     colorlinks=true,
     linkcolor=myblue,
     filecolor=magenta,      
     urlcolor=myblue,
     citecolor = myblue
}

\begin{document}

\title{Fault-Tolerant Code Switching Protocols for Near-Term Quantum Processors}

\author{Friederike Butt}
\email{friederike.butt@rwth-aachen.de}
\affiliation{Institute for Quantum Information, RWTH Aachen University, Aachen, Germany}
 \affiliation{Institute for Theoretical Nanoelectronics (PGI-2), Forschungszentrum Jülich, Jülich, Germany}
\author{Sascha Heußen}
 \affiliation{Institute for Quantum Information, RWTH Aachen University, Aachen, Germany}
 \affiliation{Institute for Theoretical Nanoelectronics (PGI-2), Forschungszentrum Jülich, Jülich, Germany}
\author{Manuel Rispler}
 \affiliation{Institute for Quantum Information, RWTH Aachen University, Aachen, Germany}
 \affiliation{Institute for Theoretical Nanoelectronics (PGI-2), Forschungszentrum Jülich, Jülich, Germany}
\author{Markus Müller}
 \affiliation{Institute for Quantum Information, RWTH Aachen University, Aachen, Germany}
 \affiliation{Institute for Theoretical Nanoelectronics (PGI-2), Forschungszentrum Jülich, Jülich, Germany}

\date{\today}
 
\begin{abstract}
\justifying

Topological color codes are widely acknowledged as promising candidates for fault-tolerant quantum computing. Neither a two-dimensional nor a three-dimensional topology, however, can provide a universal gate set \{H, T, CNOT\}, with the T-gate missing in the two-dimensional and the H-gate in the three-dimensional case. These complementary shortcomings of the isolated topologies may be overcome in a combined approach, by switching between a two- and a three-dimensional code while maintaining the logical state. 
In this work, we construct resource-optimized deterministic and non-deterministic code switching protocols for two- and three-dimensional distance-three color codes using fault-tolerant quantum circuits based on flag-qubits. Deterministic protocols allow for the fault-tolerant implementation of logical gates on an encoded quantum state, while non-deterministic protocols may be used for the fault-tolerant preparation of magic states. 
Taking the error rates of state-of-the-art trapped-ion quantum processors as a reference, we find a logical failure probability of $3\%$ for deterministic logical gates, which cannot be realized transversally in the respective code. 
By replacing the three-dimensional distance-three color code in the protocol for magic state preparation with the morphed code introduced in~\cite{vasmer2022morphing}, we reduce the logical failure rates by two orders of magnitude, thus rendering it a viable method for magic state preparation on near-term quantum processors.
Our results demonstrate that code switching enables the fault-tolerant and deterministic implementation of a universal gate set under realistic conditions, and thereby provide a practical avenue to advance universal, fault-tolerant quantum computing and enable quantum algorithms on first, error-corrected logical qubits.
\end{abstract}

\maketitle

\section{Introduction}

Universal quantum computation holds the promise to perform certain computational tasks exponentially faster than any known classical algorithm~\cite{shor1994algorithms, grover1997quantum}. In the current noisy intermediate-scale quantum (NISQ) era, however, the accuracy of quantum algorithms is limited by noise~\cite{preskill2018quantum}. 
A prospective means of increasing their robustness against the noise is to encode quantum information on logical qubits, with each logical qubit consisting of multiple physical qubits.
On these logical qubits, quantum error correction (QEC)~\citep{gottesman1997stabilizer, Nielsen_and_Chuang} can be performed to correct for errors on physical qubits and, thereby, to recover the initially encoded information~\cite{terhal2015quantum, preskill1998reliable}. 
For physical error rates below a certain threshold, QEC enables practical quantum computing for arbitrarily long times given suitable, i.e.~fault-tolerant (FT), circuit constructions ~\cite{kitaev1997quantum, aliferis2005quantum, knill1998resilient, aharonov1997fault}. 
FT circuits can be designed by using \textit{transversal} gate operations, which are composed of single-qubit unitaries acting on individual qubits in each encoded block, such that potential errors in any of these operations are prevented from proliferating uncontrollably~\cite{Nielsen_and_Chuang, mermin2007quantum}. 
In order to approximate arbitrary computations on encoded qubits, a discrete set of operations that forms a \textit{universal} gate set can be used, which requires at least one non-Clifford gate~\cite{kitaev1997quantum, solovay1995lie}. 
However, the Eastin--Knill theorem states that no QEC code exists that has a universal, transversally encoded and, therefore, FT gate set~\cite{eastin2009restrictions}. 
This complicates the implementation of a universal FT gate set and poses a key challenge towards error-corrected universal quantum computing. 

\begin{figure*}[t]
    \centering
    \includegraphics[width=0.99\textwidth]{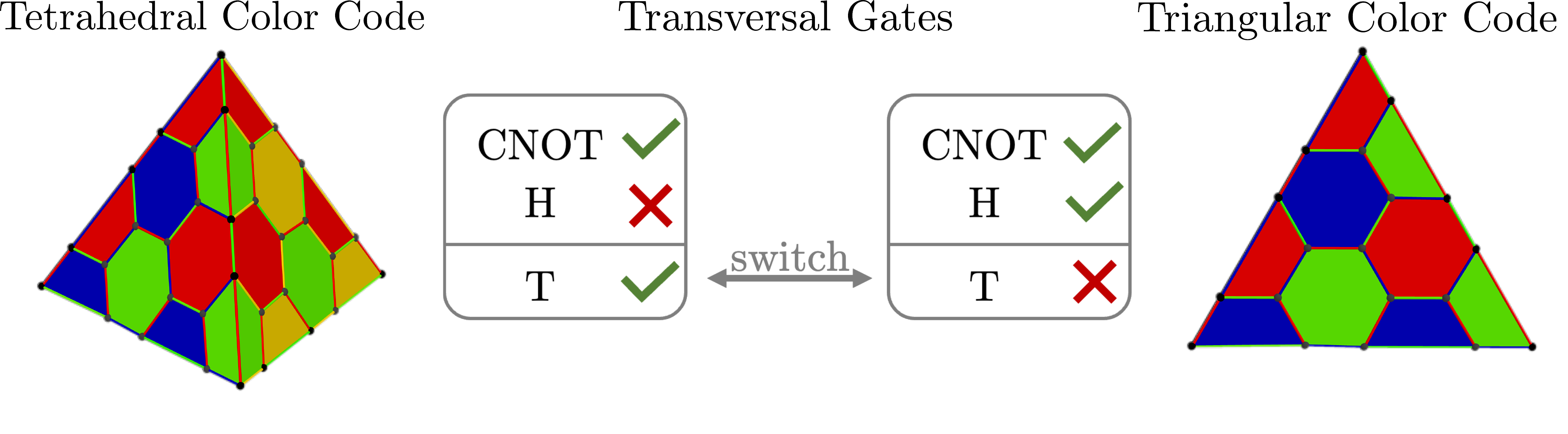}
    \caption{\justifying \textbf{Transversal gate set of three-dimensional tetrahedral and two-dimensional triangular color codes.} A two-dimensional triangular color code, as depicted on the right, has a transversal implementation of the CNOT- and the Hadamard-gate. On the left side, a three-dimensional tetrahedral code is illustrated, which has a transversal implementation of the CNOT-gate and the non-Clifford T-gate. By switching between these two code classes, it is possible to access all gates of a universal gate set \{CNOT, H, T\} with transversal implementations. 
}
    \label{fig:universal_gate_set_color_codes}
\end{figure*}

Recent quantum computing experiments are focused on the practical encoding of qubits on a logical level and investigate the implementation of a quantum memory. 
For example, on trapped-ion systems, the preparation of logical states~\cite{nigg2014quantum}, error detecting codes~\cite{linke2017fault}, FT stabilizer readouts~\cite{hilder2022fault} in a shuttling based architecture~\cite{wan2020ion}, as well as repeated cycles of QEC~\cite{ryan2021realization} have been implemented. 
In  superconducting qubits, logical qubits have successfully been initialized~\cite{takita2017experimental, satzinger2021realizing}, while repeated QEC has been realized~\cite{krinner2022realizing, google2021exponential, zhao2022realization}, as well as error detection codes~\cite{andersen2020repeated, kelly2015state, andersen2020repeated}. Recently, a distance-five surface code logical qubit outperformed a distance-three logical qubit, demonstrating an improvement of performance of QEC codes with an increasing number of qubits~\cite{google2023suppressing}. 
Rydberg atoms are a promising candidate for building quantum processors due to their strong long-range interactions and scalability~\cite{browaeys2020many, saffman2016quantum} and have shown rapid progress in single- and multi-qubit control \cite{levine2019parallel, graham2022multi, evered2023high} as well as first elements of quantum error correction~\cite{bluvstein2022quantum}. 
Advancements in other qubit architectures have been reported, as for example a three-qubit phase-correcting code  in a silicon-based architecture~\cite{takeda2022quantum} and fault-tolerant operations on a logical qubit using spin qubits in diamond~\cite{abobeih2022fault}, among others.

\begin{figure*}[tp]
    \centering
    \includegraphics[width=0.99\textwidth]{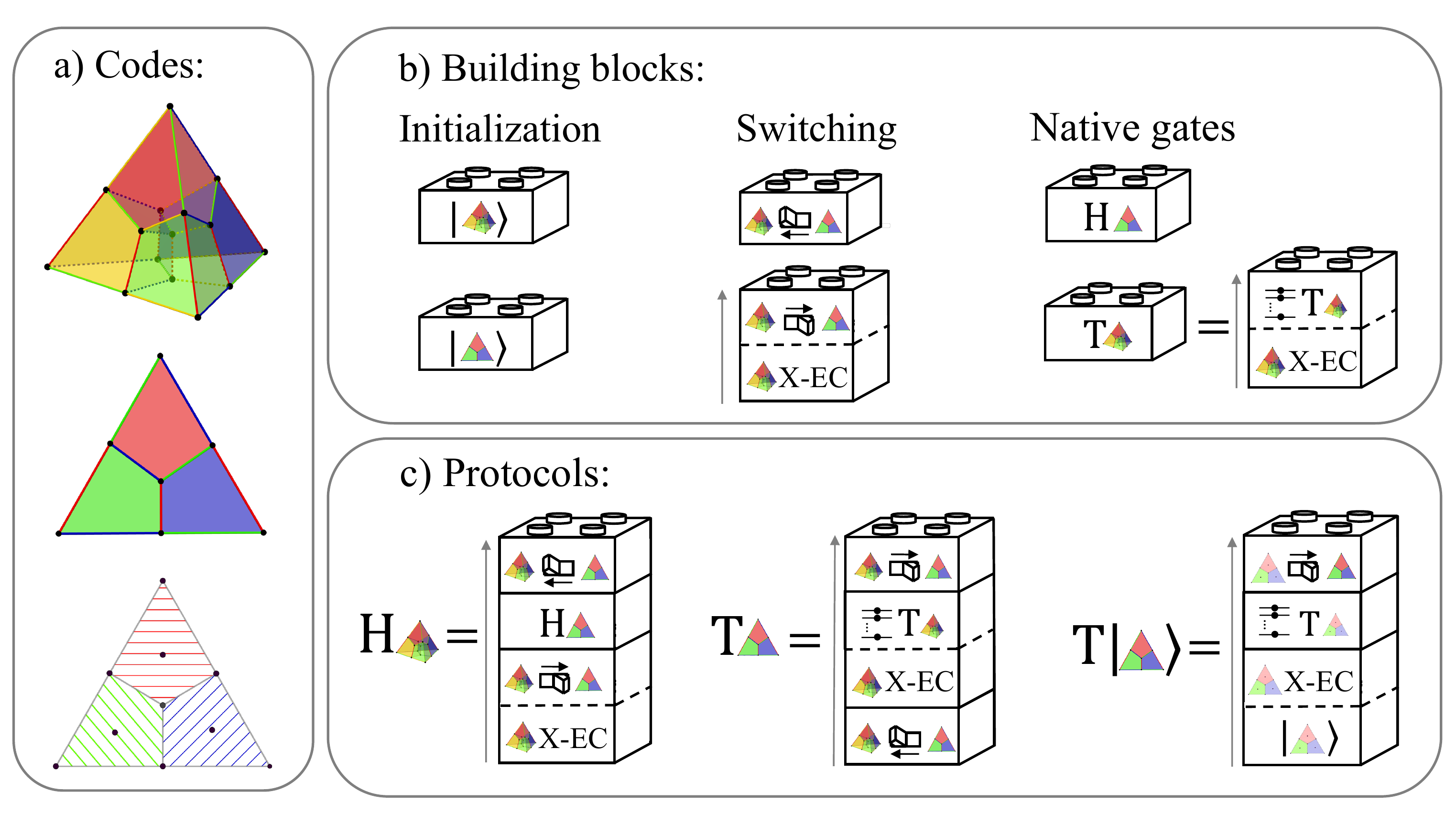}
    \caption{\justifying \textbf{Codes and FT building blocks used to compose FT code switching protocols.} (a) The tetrahedral $[[15, 1, 3]]$ code (top) contains $15$ physical qubit, encodes one logical qubit and has distance three. The two-dimensional $[[7, 1, 3]]$ Steane code (center) consists of $7$ physical qubits and has the same distance and number of logical qubits. The morphed $[[10, 1, 2]]$ code (bottom) is made up of $10$ physical qubits, encodes a single logical qubit, but is a distance-two code, so arbitrary single errors are detectable, yet not correctable. 
    When constructing code switching schemes, we distinguish between deterministic and non-deterministic protocols. Deterministic protocols are implemented using the tetrahedral code, correcting for any single error in each block. Non-deterministic protocols include postselecting during switching and replacing the tetrahedral code with the error detecting morphed $[[10, 1, 2]]$ code. 
    (b) FT building blocks for the considered codes. These include the unitary initialization of logical states on the considered codes (left). The unitary encoding circuits are given in Fig.~\ref{fig:init_tetra_logical_0_circuit}, \ref{fig:init_tetra_logical_plus_circuit}, \ref{fig:encoding_bulk} and \ref{fig:10_code_circuits}. 
    The central blocks represent FT switching to and from the $[[7, 1, 3]]$ Steane code. The \textsc{X-EC} block represents one round of X-error correction, which is required when switching to the two-dimensional code. 
    In the illustrated blocks, each tetrahedron can be replaced by the morphed $[[10, 1, 2]]$ code to obtain the corresponding block for the morphed code. 
    (c) The FT building blocks illustrated in (b) can be combined into FT protocols. The logical FT Hadamard-gate for the tetrahedral $[[15, 1, 3]]$ code (left) can be implemented by, first, switching to the Steane code, then, applying the transversal Hadamard-gate and, lastly, switching back to the tetrahedral code. Similarly, the T-gate for the $[[7, 1, 3]]$ Steane code (center) can be realized by switching to the three-dimensional code, applying the transversal T-gate and switching back to the Steane code afterwards. These two protocols correspond to deterministic FT logical operations, which are not restricted to specific input states. The right column represents the preparation of a magic state on the $[[7, 1, 3]]$ Steane code with the morphed $[[10, 1, 2]]$ code. After initializing the logical $\ket{\overline{{+}}}$-state on the morphed code, a reduced round of X-error detection is applied. 
    Then, a FT T-gate is applied to the morphed code, followed by FT switching to the $[[7, 1, 3]]$ Steane code. }
    \label{fig:protocols_overview}
\end{figure*}

These advancements in practical and scalable implementations of logical qubits enable FT operations on these encoded states and motivate the search for ways to achieve FT, universal quantum computing on near-term devices. 
The FT control of an error-corrected single logical qubit~\cite{egan2021fault} has been demonstrated on a trapped ion processor, as well as logical operations in a distance-two error-detecting surface code on a superconducting architecture~\cite{marques2022logical} and entangling gates between logical qubits~\cite{ryan2022implementing}. 
A universal set of gates was recently implemented for the first time on a seven-qubit Steane code, which is the smallest error-correcting color code, using FT circuit constructions with flag qubits~\cite{postler2022demonstration}. Here, the universal gate set is completed by using magic state injection to realize a logical non-Clifford operation. 
The non-Clifford T-gate can be implemented by preparing a magic state fault-tolerantly on an ancillary system~\cite{goto2016minimizing, chamberland2019fault} and then injecting this magic state onto the target qubits using a logical CNOT-gate~\cite{bravyi2005universal}.

An alternative to magic state injection is to complete a FT universal gate set using code switching between codes with complementary transversal gate sets~\cite{kubica2015universal, anderson2014fault, bombin2015gauge}. Code switching allows one to transfer encoded information between specific codes. To do so, stabilizers are measured to project the state onto the desired codespace. Based on the measurement outcomes, local Pauli operations are applied to change the state into the correct $+1$-eigenstate of the stabilizers of the target code. 
One candidate for implementing a universal gate set with code switching are two-dimensional color codes, because all Clifford gates can be implemented transversally~\cite{bombin2006topological, chamberland2020triangular, dehaene2003clifford, wu2023enabling}.
For a universal set of gates, at least one non-Clifford gate such as the T-gate is required,  
which can only be implemented transversally in three-dimensional codes, as for example tetrahedral color codes as illustrated in Fig.~\ref{fig:universal_gate_set_color_codes}~\cite{bombin2007topological}. 
By switching between these two- and three-dimensional color codes in a FT manner, it is possible to access all gates of a universal gate set with a transversal implementation.

To achieve fault tolerance for code switching protocols, it is not sufficient to simply use FT stabilizer measurements for error correction. The objective for syndrome measurements for error correction is to detect errors on data qubits and correct for them. 
The objective of stabilizer measurements during code switching is to project the state onto a desired codespace and apply a corresponding switching operation. 
In this case, single errors on data qubits invert the measurement outcomes and can directly introduce logical errors on the target code. Furthermore, the initial code can have different error correcting properties than the target code, which has to be carefully taken into account, as we discuss in detail in Sec.~\ref{sec:FT_CS}. 

In this work, we present strategies that allow for FT switching between two- and three-dimensional color codes. We explicitly consider the distance-three instances of these codes. These instances consist of a number of physical qubits that is available on near-term quantum processors while a single arbitrary computational error is correctable. 
We make use of ancillary flag qubits that herald the presence of errors resulting from faults in the circuits~\cite{goto2016minimizing, chao2018quantum, chamberland2019fault}. 
The concept of flag qubits has been extended to codes with arbitrary distance~\cite{ chamberland2020triangular, chamberland2018flag, tansuwannont2022achieving} and has been used for the FT initialization of logical qubits~\cite{goto2016minimizing, postler2022demonstration, chamberland2019fault} and the FT encoding of magic states~\cite{gupta2023encoding}. 
We develop FT protocols considering circuit-level noise, i.e.~noise models where all components of the underlying circuits, such as initialization, gate operations and measurements are modeled as noisy. 
We construct new flag-qubit based circuits for the encoding of logical states and for sequences of stabilizer measurements, which are resource-optimized in terms of qubit and gate count. 
Furthermore, we introduce new non-deterministic code switching protocols that make use of a transformed \textit{morphed} code, which was introduced in~\cite{vasmer2022morphing}. 
The morphed three-dimensional distance-three color code inherits the FT T-gate and makes it a candidate for completing a universal gate set. 
We construct a scheme for the FT preparation of a magic state using this new code. 
In doing so, the success rate can be increased significantly compared to the preparation of a magic state via the initial three-dimensional color code. 
Using the morphed code for magic state preparation, we find similar success rates as state-of-the-art implementations~\cite{postler2022demonstration}. 
We construct an entire toolbox of modular FT building blocks, which are illustrated in Fig.~\ref{fig:protocols_overview}. Each block is FT by itself and different blocks can be composed to FT protocols. 
Tab.~\ref{tab:resources_FT_blocks} summarizes the resources for the constructed building blocks and composite protocols, illustrated in Fig.~\ref{fig:protocols_overview}, in terms of the two-qubit-gate count and the number of qubits. 
\\

This manuscript is structured as follows. In Sec.~\ref{sec:codes_explicit} and Sec.~\ref{sec:3d_codes} we briefly review basic properties of two- and three-dimensional color codes and, specifically, the smallest instances of these code classes. In Sec.~\ref{sec:code_switching_teory}, we summarize the underlying theory for code switching. 
In Sec.~\ref{sec:FT_CS}, we discuss the different Pauli errors that have to be considered in detail and present strategies for a FT implementation of code switching. 
We introduce FT switching with the morphed code in Sec.~\ref{sec:morphing} and present results of numerical simulations of noisy circuits implementing logical operations in Sec.~\ref{sec:results} for a single-parameter noise model. In Sec.~\ref{sec:projected_performance}, we consider a multi-parameter noise model and estimate the projected performance on near-term devices, specifically focusing on state-of-the-art trapped-ion quantum processors. Lastly, we provide conclusions and an outlook in Sec.~\ref{sec:conclusion}.

\begin{table}[t]
    \centering
    \renewcommand*{\arraystretch}{1.7}
    \begin{tabular}{c||c|c}
         & \#Qubits & \#CNOT-gates\\
         \hline
         \hline
          (a) $[[15, 1, 3]] \rightarrow [[7, 1, 3]]$  & $17$ & $18$\\
          \hline
          (a) X-EC for  $[[15, 1, 3]] $  & $17$ & $120$ \\
          \hline
          (a)  $[[10, 1, 2]] \rightarrow [[7, 1, 3]]$   & $12$& $18$ \\
          \hline
          (a) X-EC for  $[[10, 1, 2]] $  & $12$ & $42$\\
          \hline
          (a) $[[7, 1, 3]]  \rightarrow  [[15, 1, 3]]$ & $17$ & $72$\\
          \hline
          (a) $[[7, 1, 3]]  \rightarrow  [[10, 1, 2]]$ & $12$& $34$ \\
          \hline
          (b) $\ket{\overline{0}}$ for $[[10, 1, 2]]$ & $11$ & $15$ \\
         \hline
          (b) $\ket{\overline{+}}$ for $[[10, 1, 2]]$ & $10$ & $14$ \\
         \hline
          (b) $\ket{\overline{0}}$ for $[[15, 1, 3]]$  & $16$ & $25$\\
          \hline
          (b) $\ket{\overline{+}}$ for $[[15, 1, 3]]$  & $16$ & $32$\\
          \hline
          (b) initialization bulk & $10$ & $20$\\
          \hline
          (c) T-gate on $[[15, 1, 3]]$ & $15$ & $0$\\
          \hline
          (c) T-gate on $[[10, 1, 2]]$ & $10$ & $6$ \\
          \hline
          (c) MS with $[[10, 1, 2]]$ & $12$ & $40$ \\
    \end{tabular}
    \caption{\justifying \textbf{Required resources for FT building blocks used for composing FT code switching protocols} as illustrated in Fig.~\ref{fig:protocols_overview}. The center column indicates the number of qubits which are required for a given block. The right column contains the number of CNOT-gates that have to be implemented for the specified protocol. It limits the circuit depth and is an indication for the experimental circuit complexity. The number of CNOT-gates is specified for the case that no error occurs during the given protocol, which in the limit of low physical error rates corresponds to the most likely case. If errors occur, this number can in- or decrease the number of required two-qubit gates. For example, if a flag qubit heralds the presence of an error, a different set of stabilizers has to be measured afterwards which changes the number of CNOT-gates. 
The number of CNOT-gates required for X-error correction on the $15$-qubit tetrahedral code is much higher than for any other block, since all of the $10$ Z-stabilizers of the tetrahedral code have to be measured twice with flags to achieve fault tolerance. }
    \label{tab:resources_FT_blocks}
\end{table}

\section{2D color codes}\label{sec:codes_explicit}

Two-dimensional topological color codes were originally proposed by Bombin et al.~\cite{bombin2006topological}. They fall into the important class of so-called CSS (Calderbank–Shor–Steane) stabilizer codes~\cite{gottesman1996class, calderbank1997quantum}. They can be specified by the code parameters $[[n, k, d]]$ where $n$ is number of physical qubits, $k$ is the number of encoded logical qubits and $d$ is the code distance. 
The simultaneous $+1$-eigenspace of all stabilizer generators corresponds to the codespace spanned by the valid logical states~\cite{kubica2015universal}. A two-dimensional color code can be constructed by placing qubits on the vertices of a three-valent and three-colorable lattice~\cite{bombin2006topological}. Different vertices are connected by \textit{links}. \textit{Plaquettes} are formed by a closed set of links. Three colors, say red, blue and green $\{R, B, G\}$ are commonly assigned to the plaquettes. Colors are assigned to links according to the color of the plaquettes they connect, so that, for example, blue plaquettes are connected by blue links. Colors are assigned in such a way that three links of different color meet at each site. By closing the boundaries periodically to form a torus, one can identify the logical operators of the two-dimensional color codes as the non-contractible loops on the torus.

A closely related class of two-dimensional color codes with open boundary conditions are triangular color codes, as exemplarily illustrated in Fig.~\ref{fig:universal_gate_set_color_codes}. 
They can be constructed by placing the described lattice on the surface of a sphere and removing one site and all its neighboring links and plaquettes. This lattice contains an odd number of physical qubits and can be deformed into a triangular shape. X- and Z-stabilizers are defined on the plaquettes of the code, which ensures that all stabilizers overlap at an even number of sites and, therefore, commute. 
Logical operators can be implemented by applying X- and Z-operations to all qubits. By applying stabilizers, one finds equivalent string-type logical operators with a minimum length $d$, which correspond to the length of one edge of the triangle. 
Triangular codes encode one logical qubit and have a transversal implementation of \{H, CNOT\}. This means that the logical $\overline{\mathrm{CNOT}}$ can be realized by applying physical CNOT-gates to pairs of qubits on two codes. The Hadamard-gate $\overline{\mathrm{H}}$ can be implemented by applying a single H-gate to each physical qubit. It interchanges $HX H = Z$ on each individual qubit and, therefore, acts as a Hadamard-gate on the logical level since the logical X- and Z-operators have the same support. It analogously maps X-stabilizers to Z-stabilizers and vice versa and, therefore, $\overline{\mathrm{H}}$ preserves the stabilizer group. 
In addition, the phase gate S can be realized transversally if one chooses a lattice tiling where the stabilizer weights are multiples of four~\cite{bombin2006topological}. With the set of gates \{H, S, CNOT\}, the whole Clifford group of gates can be generated transversally~\cite{gottesman1998heisenberg}. The smallest triangular color code with this set of transversal gates is a code that is equivalent to the Steane code~\cite{steane1996multiple}. 

\begin{figure}[t]
     \centering
     \begin{subfigure}[b]{0.22\textwidth}
         \centering
         \includegraphics[width=\textwidth]{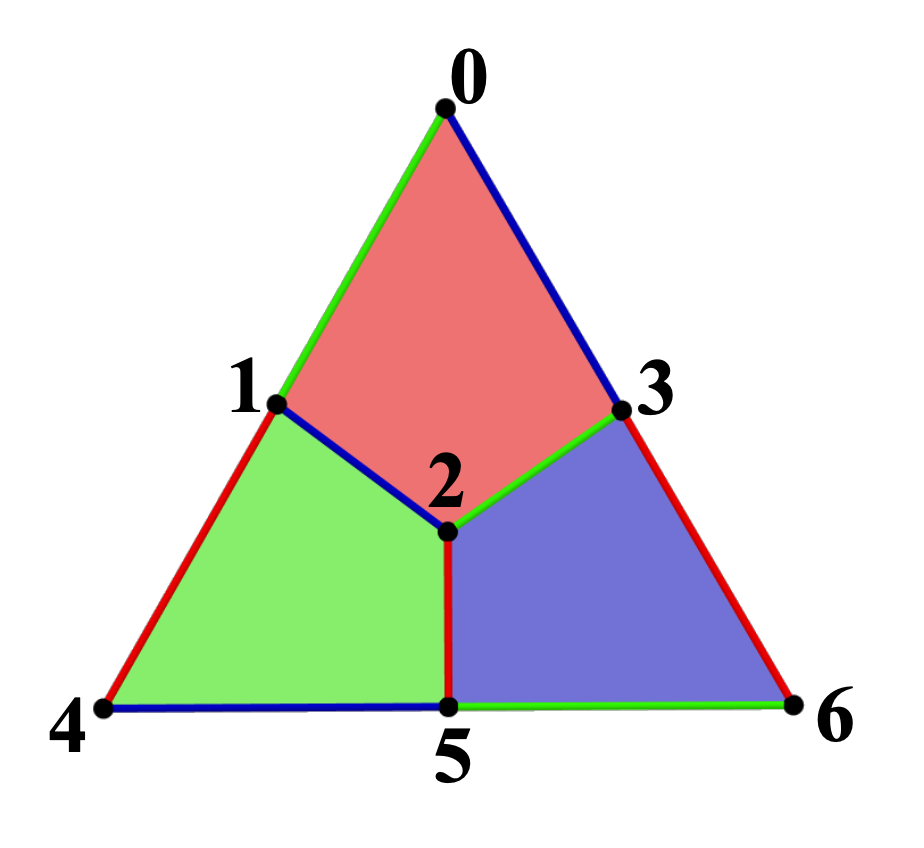}
         \caption{}
     \end{subfigure}
     \begin{subfigure}[b]{0.25\textwidth}
         \centering
         \includegraphics[width=\textwidth]{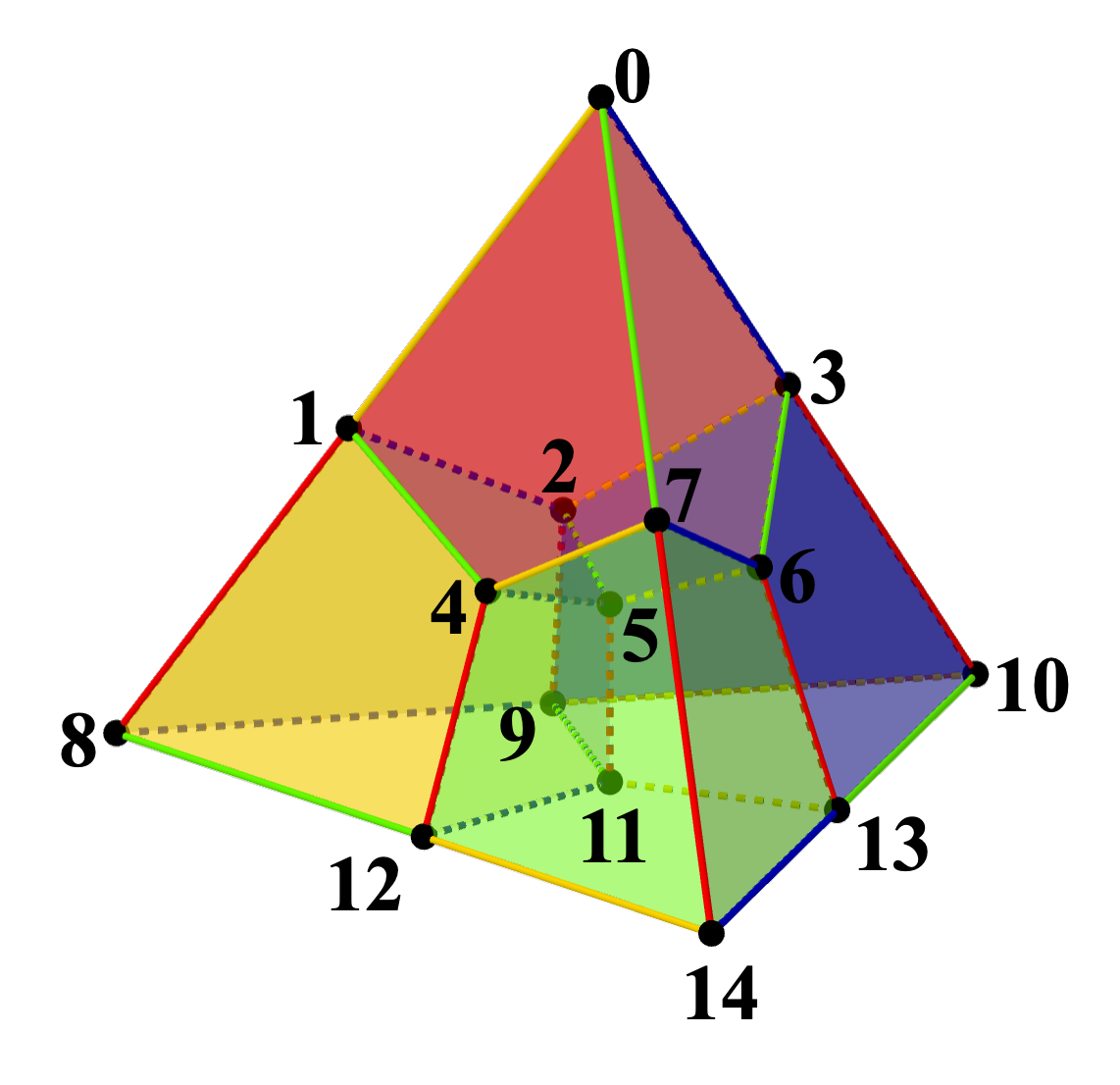}
         \caption{\centering}
     \end{subfigure}
    \caption{\justifying \textbf{Illustration of the two- and three-dimensional distance-three color codes. } (a) The $[[7, 1, 3]]$ Steane code consists of seven qubits (black dots) forming a green, red and blue plaquette (colored faces). The X- and Z-type stabilizers $S^{\sigma}_{\mathrm{col}}$ of color $ \{R, G, B\}$ are defined as $\sigma = X, Z$ Pauli operators applied to the qubits that form the given plaquette. For example, the red X-stabilizer corresponds to $S^X_R = X_0 X_1 X_2 X_3$.  The logical operators can be implemented by applying X and Z to one edge of the triangle, for example to qubits $4, 5$ and $6$. (b) The 15-qubit tetrahedral code $[[15, 1, 3]]$ has distance three, encodes one logical qubit and consists of $15$ qubits (black dots), which form three-dimensional units. These units are called cells and are labeled with the colors red, blue, green and yellow. Each cell is formed by eight qubits and shares two-dimensional boundaries (faces) with three adjacent cells. Four X-stabilizers $B^{X}_c$ are defined on the cells $c \in \{R, B, G, Y\}$ of the code, as for example the red X-stabilizer $B^X_R = X_0 X_1 X_2 X_3 X_4 X_5 X_6 X_7$. Ten independent Z-stabilizers $B^{Z}_f$ are defined on the faces $f \in \{R$, $B,$ $G,$ $Y,$ $RB$, $RG,$ $RY,$ $BG,$ $BY,$ $GY\}$ of the code, as for example the red-yellow Z-stabilizer $B^Z_{RY}= Z_1 Z_2 Z_4 Z_5$. The faces contain four qubits and can lie within the tetrahedron, or on the boundary of the tetrahedron. The logical X-operator has minimum weight $7$ and can be implemented by applying Pauli X-operators to one side of the tetrahedron, as for example the red-blue-green seven-qubit side on the right. The logical Z-operator has minimum weight $3$ and can be implemented by applying Pauli Z-operators to one edge of the tetrahedron, as for example to qubits $10$, $13$ and $14$. 
    }
    \label{fig:Steane_and_tetra_code}
\end{figure}

\subsection*{The Steane code $[[7, 1, 3]]$}
The $[[7, 1, 3]]$ Steane code consists of $n = 7$ physical qubit, encodes $k = 1$ logical qubit and has distance $d = 3$ ~\cite{steane1996multiple}. Six stabilizer generators are defined as 
\begin{align}
    S^{X}_R = X_0 X_1 X_2 X_3,  \quad S^{Z}_R = Z_0 Z_1 Z_2 Z_3 \nonumber\\
    S^{X}_G = X_1 X_2 X_4 X_5,  \quad S^{Z}_G = Z_1 Z_2 Z_4 Z_5 \\
    S^{X}_B = X_2 X_3 X_5 X_6,  \quad S^{Z}_B = Z_2 Z_3 Z_5 Z_6 \nonumber
\end{align}
with the color labels red ($R$), green ($G$) and blue ($B$) and for the indexing given in Fig.~\ref{fig:Steane_and_tetra_code}(a). The two logical operators can be implemented by applying Pauli X- and Z-operators to all seven qubit and are stabilizer-equivalent to operators of minimum weight $3$. For example, the logical Pauli-operators can be implemented on the edges of the triangle with
\begin{align}
    \overline{X} =  X_4 X_5 X_6 \quad \mathrm{and} \quad
    \overline{Z} =  Z_4 Z_5 Z_6. 
\end{align}

\section{3D color codes}\label{sec:3d_codes}

To construct a three-dimensional color code, qubits are arranged on a three-dimensional, four-valent and four-colorable lattice structure. Four colors red, green, blue and yellow \{$R, G, B, Y$\} are assigned to the smaller three-dimensional units, which are called \textit{cells}. Cells of the same color never touch but are connected by links of that same color, for example blue cells are connected by blue links~\cite{bombin2007topological}. So, a link has color $\kappa \in \{R, B, G, Y\}$ if the three cells, of which the link is part of, have colors different from $\kappa$. The two-dimensional boundaries of the cells are called \textit{faces} and have the color-labels $\kappa_1 \kappa_2$ according to the colors of the two cells of which they are part of~\cite{bombin2018transversal}. This construction can be generalized also to $n$-dimensional codes~\cite{anderson2014fault, kubica2015universal}. 

One class of three-dimensional color codes are tetrahedral codes, as exemplarily illustrated in Fig.~\ref{fig:universal_gate_set_color_codes}. Analogously to the two-dimensional case, a tetrahedral code can be constructed by arranging a three-dimensional lattice within the volume of a 3-sphere, in the above described manner. By removing one site and its neighboring cells, faces and links, the lattice contains an odd number of physical qubits and can be deformed into a tetrahedron, encoding a single logical qubit~\cite{bombin2007topological}. 
One associates faces $F$ with Z-stabilizers and cells $C$ with X-stabilizers, ensuring commutativity since each cell and face contains an even number of qubits by construction. 
A codestate $\ket{\overline{\psi}}$ of this system is characterized by the conditions
\begin{align}
    S^X_c \ket{\overline{\psi}} &= \ket{\overline{\psi}} \quad \forall c \in C, \\
    S^Z_f \ket{\overline{\psi}} &= \ket{\overline{\psi}} \quad \forall f \in F. 
\end{align}
An operator basis is defined for the qubit encoded in the tetrahedral code by
\begin{align}
    \overline{X} = X^{\otimes n} \quad \mathrm{and} \quad \overline{Z} = Z^{\otimes n}
\end{align}
with the total number of qubits $n$. 
This definition ensures that all stabilizers commute with the logical Pauli-operators, since the tetrahedral code contains an odd number of physical qubits and each stabilizer has support on an even number of sites. The $\overline{\mathrm{T}}$-gate 
\begin{align}
	\overline{\mathrm{T}} = \begin{pmatrix}
    1 & 0 \\
    0 & e^{i\frac{\pi}{4}} 
    \end{pmatrix}. 
\end{align}
can be implemented transversally on specific tetrahedral codes if the following two conditions are fulfilled. Let $M$ be subset of the set of all physical qubits $Q$. Then let T-gates be applied to the subset $M$ and T$^{\dagger}$-gates applied to the subset $M^C = Q \setminus M$. Then for the support of the X-stabilizer $G$, the transversal $\overline{\mathrm{T}}$-gate can be implemented in this manner if~\cite{bombin2018transversal, kubica2018abcs, kubica2015universal}
\begin{enumerate}\label{list:T_gate_conditions}
    \item $|M \cap G| - |M^C \cap G| = 0\,\mathrm{mod}\,8$. \\ In words, an equal number of T- and T$^{\dag}$-gates has to be applied in each cell (mod$8$). 
    \item $|M| - |M^C| = 1\,\mathrm{mod}\,8$. \\ In words, T has to be applied to one more qubit than T$^{\dag}$ in total (mod8). 
\end{enumerate}
It has been shown that the tetrahedral $[[15, 1, 3]]$ code is the smallest distance-three QEC code with a transversal non-Clifford gate~\cite{koutsioumpas2022smallest}.

\subsection*{The 15-qubit tetrahedral code $[[15, 1, 3]]$}
\begin{figure}[t]
     \centering
     \includegraphics[width=0.25\textwidth]{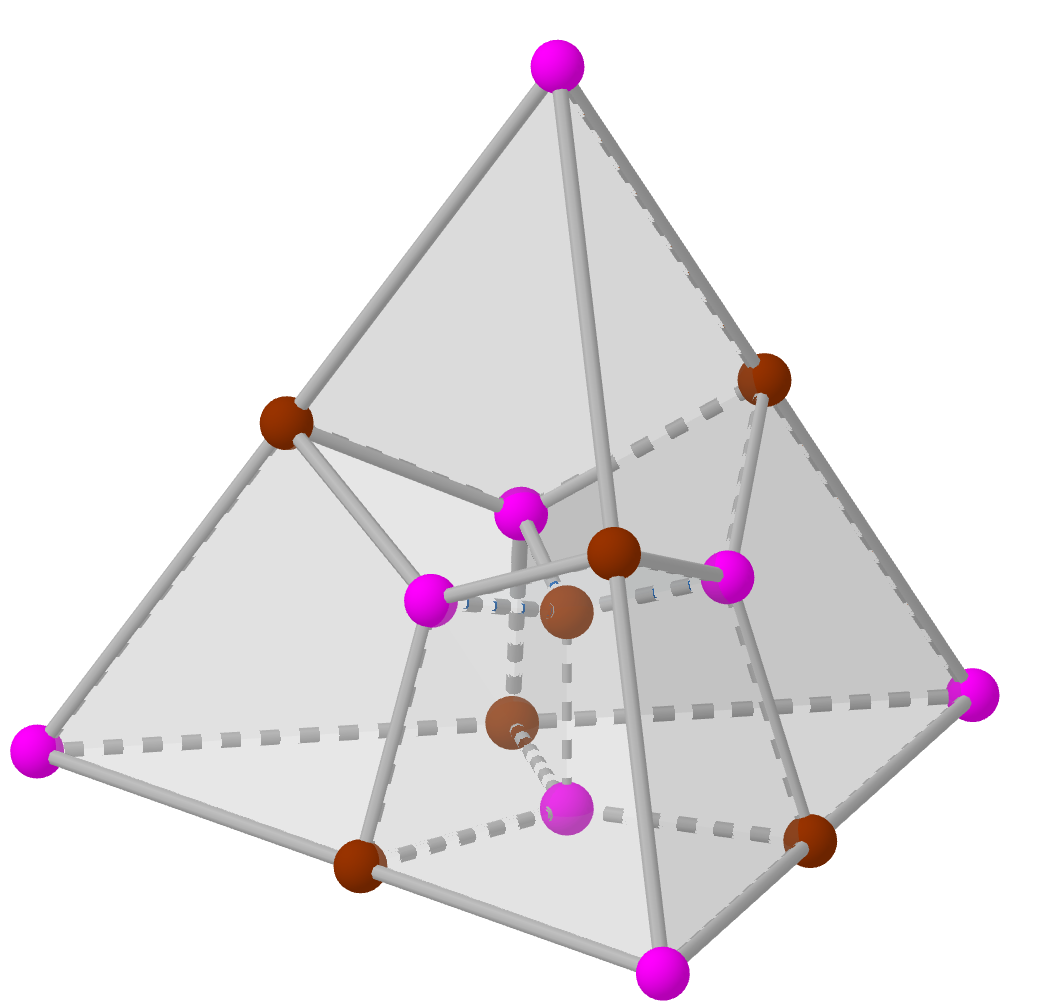}
    \caption{\justifying \textbf{Bipartition for the implementation of a transversal T-gate on the tetrahedral $[[15, 1, 3]]$ code. } The vertices in the $[[15, 1, 3]]$ code can be partitioned into two sets, so that vertices connected by an edge belong to different sets~\cite{bombin2018transversal}. Given this kind of bipartition, a $\overline{\mathrm{T}}$-gate can be implemented transversally by applying single T-gates to the qubits on the pink positions and T$^{\dag}$ to qubits on the brown positions. }
    \label{fig:T_gate_tetra}
\end{figure}

The $15$-qubit tetrahedral code $[[15, 1, 3]]$, as illustrated in Fig.~\ref{fig:Steane_and_tetra_code}(b), consists of $15$ physical qubits, one encoded logical qubit and can correct for one arbitrary error~\cite{kubica2015universal}. Four X-type stabilizers $B^{X}_c$ are defined on the cells
\begin{align}
	c \in \{R, B, G, Y\} \label{eq:X_stabs_tetra}
\end{align}
of the code, as depicted in Fig.~\ref{fig:Steane_and_tetra_code}(b). Ten independent Z-type stabilizers $B^{Z}_f$ are defined on the faces 
\begin{align}
	f \in \{R, B, G, Y, RB, RG, RY, BG, BY, GY\}  \label{eq:Z_stabs_tetra}
\end{align}
of the code. The logical operators are defined as
\begin{align}
    \overline{X} = X^{\otimes 15} \quad \mathrm{and} \quad \overline{Z} = Z^{\otimes 15}. 
\end{align}
These are stabilizer-equivalent to logical X-operators with minimum weight $7$ and logical Z-operators with minimum weight $3$. Accordingly, the tetrahedral $[[15, 1, 3]]$ code has distance $d_x = 7$ for X-errors and distance $d_z = 3$ for Z-errors, so it is possible to correct X-errors of weight less or equal to $3$ and any single Z-error, as well as error configurations that are stabilizer-equivalent to these. 
The $\overline{\mathrm{CNOT}}$-gate can be realized in a transversal manner by coupling two code blocks pairwise with physical CNOT gates. The H-gate cannot be implemented transversally (note that the stabilizer group is in particular not preserved under the operation $H^{\otimes n}$). In Fig.~\ref{fig:T_gate_tetra}, we show a bipartition that implements a transversal T-gate according to the above stated conditions. 

In the next section, we review how to switch between general two- and three-dimensional color codes and, specifically between the $[[7, 1, 3]]$ Steane code and the tetrahedral $[[15, 1, 3]]$ code.

\section{Code switching}\label{sec:code_switching_teory}

Encoded information can be transferred between specific stabilizer codes, if they correspond to two variants of the same subsystem code (subsystem codes also known as gauge codes). 
A subsystem code is defined by its gauge group $\mathcal{G}$, describing a general subgroup of the $n$-qubit Pauli group on a set of physical qubits~\cite{poulin2005stabilizer, kribs2005unified}. The stabilizer group $\mathcal{S} \subseteq \mathcal{G}$ is the center of $\mathcal{G}$, which is generated by those Pauli operators in $\mathcal{G}$ that commute with all elements in $\mathcal{G}$. A stabilizer code can be viewed as a special case of a subsystem code where $\mathcal{S} = \mathcal{G}$. The subspace of codestates can be split into a tensor product of the logical qubits and so-called gauge qubits~\cite{kubica2015universal}
\begin{align}
    \ket{\psi} = \ket{\overline{\psi}} \otimes \ket{g}_G. 
    \label{eq:subsystem_state}
\end{align}
Here, $\ket{\overline{\psi}}$ represents the logical state and $\ket{g}_G$ represents the gauge state. The gauge state corresponds to extra degrees of freedom which are not uniquely fixed by the stabilizers. An element $U$ of $\mathcal{G} \setminus \mathcal{S}$ has the effect on this state $\ket{\psi}$
\begin{align}
    U \ket{\psi} = \ket{\overline{\psi}} \otimes \left( U \ket{g}_G \right),\label{eq:gauge_fix} 
\end{align}
leaving the logical state $\ket{\overline{\psi}}$ unchanged and only affecting the gauge state $\ket{g}_G$. Thereby, equivalent codestates can be generated by applying elements of $\mathcal{G} \setminus \mathcal{S}$ without changing the encoded information. A logical gate $\overline{L}$ can affect both the logical state and the gauge state, while preserving the codespace~\cite{kubica2015universal}
\begin{align}
    \overline{L} \ket{\psi} = \left( \overline{L'}\ket{\overline{\psi}} \right) \otimes  \ket{g'}_G . 
\end{align}\label{eq:dressed_gate}

\begin{figure}[t]
     \centering
     \includegraphics[width=0.4\textwidth]{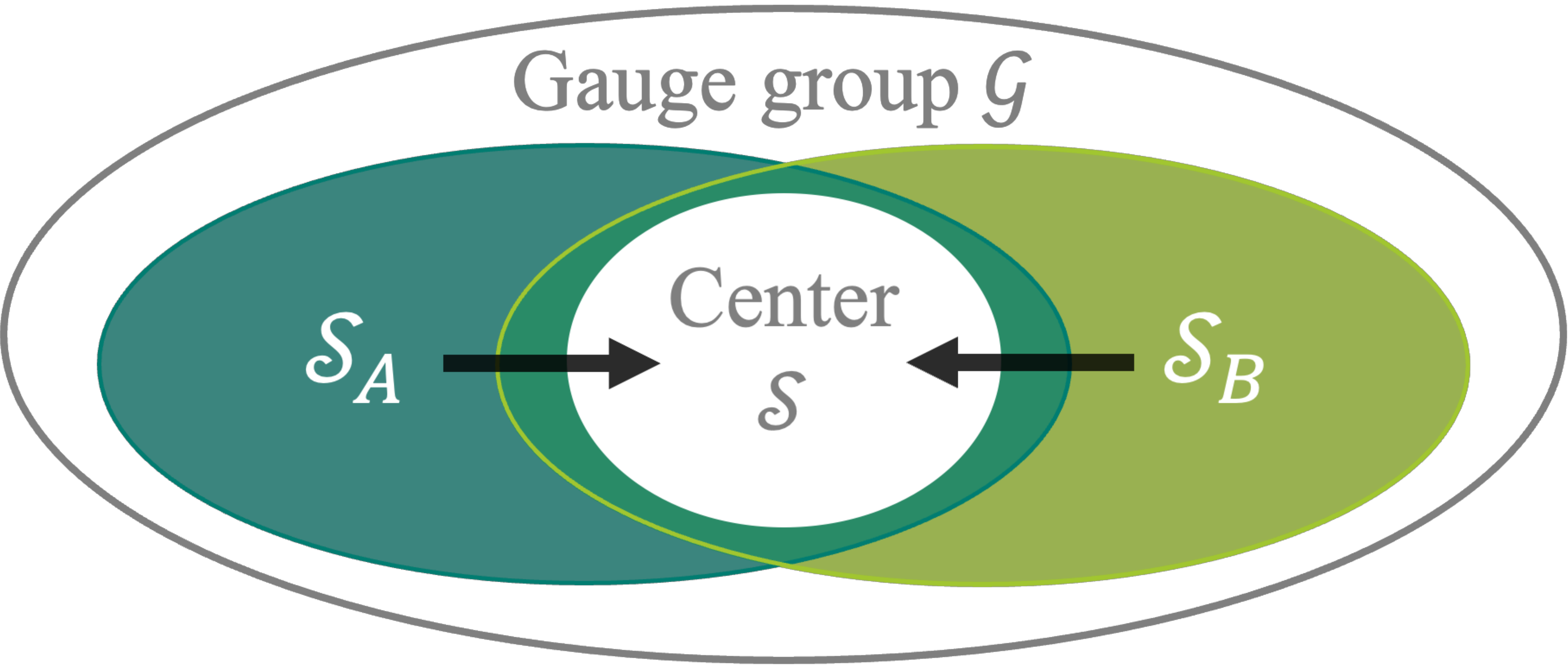}
    \caption{\justifying \textbf{Illustration of the stabilizer groups of two stabilizer codes $A$ and $B$ within the same subsystem code. } The stabilizer groups of a stabilizer code $\mathcal{S}_A$ and $\mathcal{S}_B$ are part of the gauge group of the corresponding subsystem code $\mathcal{G}$ with its center $\mathcal{S}$. Both codes $\mathcal{S}_A$ and $\mathcal{S}_B$ contain $\mathcal{S}$. Therefore, codestates in $A$ and $B$ can be written as tensor products in the subsystem $\ket{\psi}_A = \ket{\overline{\psi}} \otimes \ket{g_A}_G$ and $\ket{\psi}_B = \ket{\overline{\psi^{'}}} \otimes \ket{g_B}_G$. }
    \label{fig:subsystemillustration}
\end{figure}

Given a stabilizer code $\mathcal{S}_A$, for which $\mathcal{S} \subset \mathcal{S}_A$, any $+1$-eigenstate of $\mathcal{S}_A$ is consequentially also a codestate in the subsystem $\mathcal{S}$. Therefore, any codestate $\ket{\psi}_A \in \mathcal{S}_A$ is also a subsystem codestate, i.e. it can be written as a tensor product of a logical state and a gauge state as in Eq.~\eqref{eq:subsystem_state}
\begin{align}
    \ket{\psi}_A = \ket{\overline{\psi}} \otimes \ket{g_A}_G. 
\end{align}
Analogously, a second stabilizer code $\mathcal{S}_B$ can be defined, for which $\mathcal{S} \subset \mathcal{S}_B$, so we can write 
\begin{align}
	\ket{\psi}_B = \ket{\overline{\psi^{'}}} \otimes \ket{g_B}_G. 
\end{align}
The stabilizer groups of the three codes considered are illustrated in Fig.~\ref{fig:subsystemillustration}. If the logical operators $\overline{L}$, $\overline{L}_A$ and $\overline{L}_B$ of the three codes can be represented in the same way, and $\ket{\psi}_A$ and $\ket{\psi}_B$ correspond to the same logical state in their stabilizer code, they must be logically equivalent in the subsystem code. This means that $\ket{\overline{\psi}}$ and $\ket{\overline{\psi'}}$ are the same logical state for both 
\begin{align}
    \ket{\psi}_A &= \ket{\overline{\psi}} \otimes \ket{g_a}_G \quad \mathrm{and} \nonumber \\ \ket{\psi}_B &= \ket{\overline{\psi'}} \otimes \ket{g_b}_G. \label{eq:tensor_product}
\end{align}
So, in the subsystem code, $\ket{\psi}_A$ and $\ket{\psi}_B$ only differ in their gauge state $\ket{g_A}$, $\ket{g_B}$. Elements of $\mathcal{G} \setminus \mathcal{S}$ only affect the gauge state and leave the logical state unchanged. These elements can be used to \textit{fix} the gauge state from $\ket{g_A}$ to $\ket{g_B}$ or vice versa, while leaving the logical state unchanged. 

Now, let us consider switching from code $A$ to code $B$. First, those stabilizers of the target code $B$, which are not fulfilled initially, have to be measured. This measurement projects the state randomly into a $\pm1$-eigenstate of the measured stabilizer. Secondly, elements of $\mathcal{G} \setminus \mathcal{S}$ are applied, which only affect the gauge state, to force the gauge state into the corresponding state $\ket{\psi}_B = \ket{\overline{\psi}} \otimes \ket{g_B}_G$. 

\subsection{Code switching between the $[[7, 1, 3]]$ and $[[15, 1, 3]]$ }
Switching between the Steane code and the $15$-qubit tetrahedral code has been discussed and analyzed in~\cite{anderson2014fault, kubica2018abcs} and~\cite{beverland2021cost}, among others. In the following, 
we briefly review Non-FT switching between these codes before presenting a FT scheme in Sec.~\ref{sec:FT_CS}. 

One can define a subsystem code whose gauge group $\mathcal{G}$ is generated by all independent X- and Z-faces that can be defined on the tetrahedral structure.  The stabilizer group of this subsystem code $\mathcal{S}_{\mathrm{subsystem}}$ is generated by those elements that commute with all other elements of $\mathcal{G}$ and is therefore generated by the X- and Z-cells of the tetrahedron. In contrast to the stabilizers of the stabilizer code $[[15, 1, 3 ]]$, the Z-stabilizers of the subsystem code are not defined on the ten independent faces of the code, but only on the four weight-$8$ cells. The stabilizer group of the subsystem is part of both the stabilizer group of the Steane code, together with the bulk, and of the stabilizer group of the $[[15, 1, 3]]$ tetrahedral code, meaning 
\begin{align}
    \mathcal{S}_{\mathrm{subsystem}} &\subset (\mathcal{S}_{\mathrm{Steane}}, \mathcal{S}_{\mathrm{bulk}}), \nonumber\\
    \mathcal{S}_{\mathrm{subsystem}} &\subset \mathcal{S}_{\mathrm{tetrahedron}}. 
\end{align}

Here, the bulk is formed by those data qubits of the tetrahedron that are not part of the Steane code (yellow cell). 
Note that, if the state fulfills two opposing faces individually, then it is also a $+1$-eigenstate of its products, which corresponds to the composed cell. Consequentially, every codestate of the tetrahedral stabilizer code, which is a $+1$-eigenstate of all elements of $\mathcal{S}_{\mathrm{tetra}}$, is always also a codestate of the subsystem code. Analogously, every codestate of the Steane code together with the bulk is also always a codestate of the subsystem. So, one can write codestates of the $[[15, 1, 3]]$ code and the $[[7, 1, 3]]$ Steane code in the subsystem as a tensor product of the corresponding logical state and a specific gauge state, as given in Eq.~\eqref{eq:tensor_product}. 

If the logical operators of the three codes have a common representation, so that $\ket{\overline{\psi}}$ is the same logical state for both codes, they must be logically equivalent in the subsystem code. One common representation for the logical operators of the Steane code and the $[[15, 1, 3]]$ tetrahedral code corresponds to one side of the tetrahedron, such as 
\begin{align}
    \overline{X} &= X_0 X_3 X_6 X_7 X_{10} X_{13} X_{14} \quad \mathrm{and} \nonumber \\ 
    \quad \overline{Z} &= Z_0 Z_3 Z_6 Z_7 Z_{10} Z_{13} Z_{14}
\end{align}
for the indexing given in Fig.~\ref{fig:Steane_and_tetra_code}(b). 
To switch between the two codes $[[7, 1, 3]]$ and $[[15, 1, 3]]$, first, stabilizers of the target code have to be measured. This projects the encoded state into a $+1$- or $-1$-eigenstate of the measured stabilizers. For a negative measurement outcome, a combination of gauge-operators has to be applied that forces the state into a $+1$-eigenstate of all stabilizers of the target code. 

Concretely, to switch from the $[[15, 1, 3]]$ code to the Steane code, first, one has to measure the three X-plaquettes of the Steane code, which are $S_R^X = X_0 X_3 X_6 X_7$, $S_B^X = X_3 X_6 X_{10} X_{13}$ and $S_G^X = X_6 X_7 X_{13} X_{14}$ for the indexing given in Fig.~\ref{fig:Steane_and_tetra_code}(b). Based on these measurement outcomes, a combination of Z-faces connecting the Steane code with the bulk is applied in order to force the state into the codespace of the Steane code. For example, let us consider the case where the measurement outcome of the X-plaquettes of the Steane code is $(S^X_R$, $S^X_B$, $S^X_G) = (1, 0, 0) $. This means that the measurement projected the state onto a $+1$-eigenstate of the blue and green plaquettes and a $-1$-eigenstate of the red plaquette. In order to force the state into the desired codespace, we apply the gauge operator $B_{BG}^Z = Z_5 Z_6 Z_{13} Z_{14}$, as illustrated in Fig.~\ref{fig:dangerous_errors_switching}(a). This operator overlaps at an even number of sites with the blue and green plaquette, so the state stays a $+1$-eigenstate of these stabilizers. It overlaps at a single site with the red plaquette and, therefore, forces the state into a $+1$-eigenstate of this plaquette. 
Note that this is in contrast to using single Pauli-operations as one would do for correcting errors on a set of qubits. This would affect both the gauge state $|g\rangle_G$ and the logical state $|\overline{\psi}\rangle$, as given in Eq.~\ref{eq:subsystem_state} and, therefore, change the encoded information. 

This protocol is inverted for the other direction. To switch from the Steane code to the $[[15, 1, 3]]$ code, first, one has to measure the three Z-faces that connect the Steane code and the bulk, which are $B_{RB}^Z = Z_2 Z_3 Z_5 Z_6$, $B_{RG}^Z = Z_4 Z_5 Z_6 Z_7$ and $B_{BG}^Z = Z_5 Z_6 Z_{11} Z_{13}$. Based on these measurement outcomes, a combination of the Steane X-plaquettes is applied in order to change the state into the codespace of the $[[15, 1, 3]]$ code. 

\section{Deterministic FT code switching}\label{sec:FT_CS}

For the physical implementation of code switching protocols on quantum processors, the underlying operations have to be constructed in a way that is resilient against noise. In this section, we aim at finding schemes that tolerate any single error on an arbitrary component of the circuit. Then, the probability that two components of the circuit fail at the same time and thus cause a logical failure is order $p^2$, where $p$ is the failure probability for a single component~\cite{Nielsen_and_Chuang}. 
For deterministic schemes, we should be able to correct for any single error reliably. In order to construct these FT code switching protocols, a set of criteria has to be fulfilled. 

The first criterion is that the error configuration on the initial code also has to be correctable on the target code. 
For the color codes considered, it is possible to correct weight-three X-errors on the three-dimensional code, while it is only possible to correct a single X-error on the two-dimensional code. Starting in a codestate of the three-dimensional code that contains a weight-three error configuration, this would initially be correctable. Assuming perfect switching, which does not introduce any additional errors, this error configuration would be transferred onto the two-dimensional code, where a weight-three error configuration can directly correspond to a logical operator. Therefore, we perform one round of X-error correction with flag qubits before switching to the two-dimensional code. This corrects any previously present configuration with up to three X-errors and corrects for any single error that occurs during the EC procedure. 

\begin{figure*}[t]
     \centering
	\includegraphics[width=\textwidth]{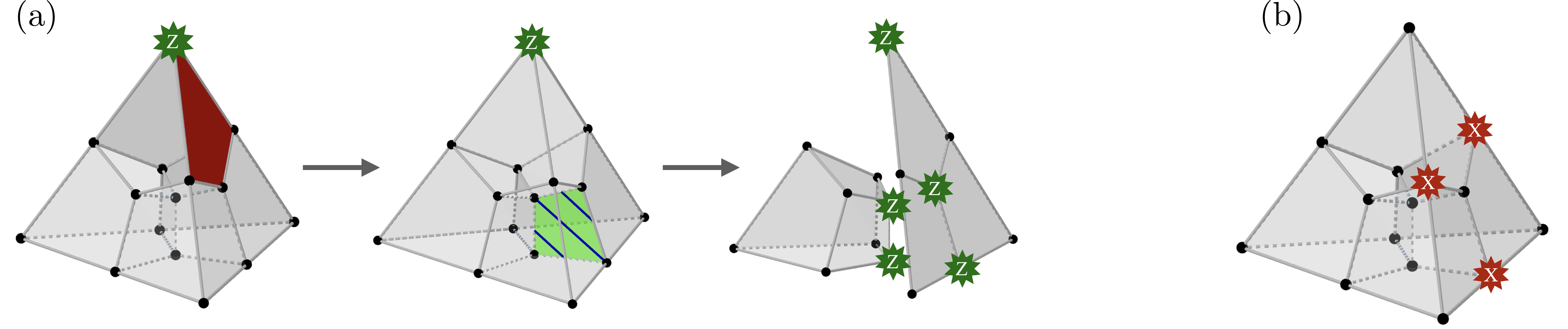}
    \caption{\justifying \textbf{Error configurations on data qubits that can lead to a logical failure. } (a) Consider switching from the $15$-qubits tetrahedral code to the target Steane code, which lives on the seven qubits forming the right three-plaquette triangle of the tetrahedron. To transfer the encoded information to the Steane code, we measure the three X-plaquettes of the Steane code. If for example a Z-error has occurred on the top qubit of the red cell (large green position on the left) before the measurement, this inverts the measurement outcome of the corresponding plaquette. If we originally had measured the syndrome $(S^X_R$, $S^X_B$, $S^X_G)$ $=$ $(0, 0, 0) $, we would now find $(S^X_R, S^X_B, S^X_G)$ $= (1, 0, 0) $. Based on this measurement outcome, we would apply $B^Z_{BG}$ to switch to the Steane code (center). If we then consider the Steane code (right), we effectively applied Pauli Z-operations to the three qubits on the large green positions. This directly corresponds to a logical $\overline{Z}$ on the Steane code and results in a logical failure. (b) The three positions where a single X-error leads to a logical failure when switching from the Steane code to the tetrahedral code are depicted in red. }
    \label{fig:dangerous_errors_switching}
\end{figure*}

The second condition is that we have to be able to identify errors on data qubits. Specific errors on data qubits invert the outcome of the stabilizer measurements. If we cannot distinguish this inverted measurement outcome from the original information, we do not notice that an error has occurred and a logical error can be introduced, as illustrated in Fig.~\ref{fig:dangerous_errors_switching}. Therefore, one does not directly achieve fault tolerance for code switching by simply using the concept of FT stabilizer measurements with flag qubits. Considering realistic circuitry with noisy components, it is necessary to consider these errors on data qubits as well. 

\begin{figure*}[t]
     \centering
	\includegraphics[width=\textwidth]{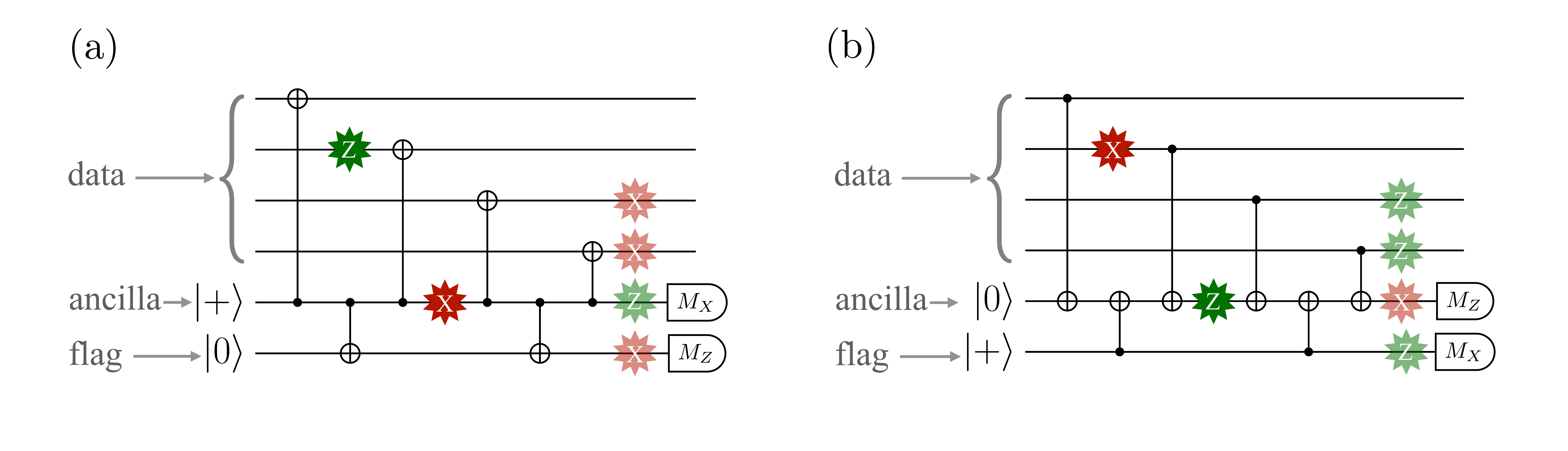}
    \caption{\justifying \textbf{Circuits for stabilizer measurements using flag qubits~\cite{chao2018quantum}.} (a) Circuit for measuring a weight-$4$ Pauli X-operator containing an ancilla qubit which is coupled to the data qubits that participate in the corresponding measurement. An additional flag qubit is coupled with CNOT-gates to the first ancilla. An X-error in the middle of the circuit, as depicted in red, propagates onto two data qubits and the flag qubit. Measuring the flag qubit in the end indicates that a potentially dangerous error has occurred. By performing an additional Z-face stabilizer measurement without flags afterwards, one can detect these errors and correct for them. Since an error before the second or after the next to last CNOT-gate only results in an error which is equivalent to a weight-$1$ error on the data qubits, this scheme is FT with respect to errors on the ancilla qubits. A Z-error during this X-plaquette measurement, as depicted in green, propagates onto the ancilla qubit, which changes the measured syndrome as illustrated in Fig.~\ref{fig:dangerous_errors_switching}. In order to avoid applying a logical error, these errors on the data qubits have to be detected and corrected. (c) Circuit for measuring a weight-$4$ Pauli Z operator using an additional flag qubit. Analogously, Z-errors on the ancilla qubits are detected by the flag qubit and can be corrected by performing an additional stabilizer measurement. X-errors on the data qubits during the Z-face measurement can propagate and cause a logical failure. }
    \label{fig:flag_measurements_circuits}
\end{figure*}

In general, we can obtain the expectation value of an operator $\overline{\mathrm{O}}$, that has a transversal implementation on a given code, by coupling an ancilla qubit to the data qubits that participate in the measurement~\cite{preskill1998fault}. 
Using this kind of circuit to extract the syndrome is, however, not FT because single faults can directly cause a logical failure. A single fault on 

\begin{minipage}[b]{1\linewidth}
	\begin{enumerate}[label=(\Alph*)]
		\item ancilla qubits during a stabilizer measurement, 
		\item data qubits before or during a stabilizer measurement, 
		\item a measurement of the ancilla qubit at the end of a stabilizer readout
	\end{enumerate}
\end{minipage}

\noindent can induce a logical error, which we discuss in detail in the remainder of this section.

\subsection{Errors on ancilla qubits}\label{sec:errors_ancillas}

Single errors on the ancilla qubit can propagate onto data qubits. An X-error after the second CNOT-gate in the circuit shown in Fig.~\ref{fig:flag_measurements_circuits}(a) propagates onto two data qubits. On the target Steane code, this results in a logical failure, since in the Steane code, only a single X-error is correctable. Analogously, Z-errors on the ancilla qubit during the measurement of a Z-stabilizer can lead to a logical error $\overline{\mathrm{Z}}$ on the target code, since only a single Z-error is correctable in the $[[15, 1, 3]]$ tetrahedral stabilizer code. 
We can implement stabilizer measurements, which are FT with respect to these errors on ancillary qubits, using a \textit{flag}-based measurement scheme, as depicted in Fig.~\ref{fig:flag_measurements_circuits}. The concept of flag qubits was introduced in~\cite{chao2018quantum} and has, for example, been used for FT error correction in color codes in two and three dimensions~\cite{chamberland2020triangular, chamberland2018flag, tansuwannont2022achieving} and the initialization of logical qubits~\cite{goto2016minimizing, postler2022demonstration, chamberland2019fault}. Again, an ancilla qubit is coupled to the data qubits via CNOT-gates and a flag qubit is coupled to the ancilla, as depicted in Fig.~\ref{fig:flag_measurements_circuits}. If a dangerous fault occurs on the ancilla qubit during this measurement, it still propagates onto one or more of the data qubits, but it also propagates onto the flag qubit. In that case, the flag qubit will be measured in the $-1$-state in the end and indicates that a potentially dangerous error has occurred.

If a circuit flags, meaning that a flag qubit was measured in the corresponding $-1$-eigenstate, we still have to identify which error has occurred. The circuit could have flagged owing to a measurement error in the end, or to an error before the second to last CNOT-gate. Specifically, we have to check if the flag error leads to a logical failure. Here, these dangerous flag errors correspond to propagated weight-$2$ errors on the data qubits resulting from a single error on the ancilla qubit, as depicted in Fig.~\ref{fig:flag_measurements_circuits}. In order to localize these errors, one additional Z- (X-) stabilizer has to be measured. 
After this additional measurement of a single stabilizer, the complete syndrome is measured again without flag qubits. 
Together with the information which circuit flagged before, the dangerous weight-2 errors can be identified and corrected. 

\subsection{Errors on data qubits}\label{sec:errors_data}

Errors on the data qubits can propagate onto the ancilla qubit and invert the measurement outcome, as illustrated in Fig.~\ref{fig:flag_measurements_circuits}. Taking this inverted measurement outcome would cause us to apply the incorrect switching operation and can lead to a logical failure on the target code. 

For example, a single Z-error on the top qubit $0$ would invert the measurement outcome of the red plaquette, when switching from the tetrahedral code to the Steane code. We would then apply the according switching operation, as illustrated in Fig.~\ref{fig:dangerous_errors_switching}(a). But owing to the inverted measurement outcome, we now introduce additional errors on the data qubits of the Steane code that overlap with the applied gauge operator. Together with the initial error, this can directly correspond to a logical operator on the Steane code. For FT code switching, these errors on data qubits have to be corrected. 

To identify and correct for these errors on data qubits, one can exploit that the state should always be a $+1$-eigenstate of the complete cells. This means that, for switching from the $[[15, 1, 3]]$ code to the $[[7, 1, 3]]$ Steane code, the X-faces opposite of the measured Steane plaquettes have to agree with the measured Steane X-plaquette to also fulfill the cells. For example, considering the red cell, as illustrated in Fig.~\ref{fig:Steane_and_tetra_code}, the state is either in a $+1$-eigenstate of both $S^X_R = X_0 X_3 X_6 X_7$ and $B^X_{RY} = X_1 X_2 X_4 X_5$ or in a $-1$-eigenstate of both these operators so that the state is a $+1$-eigenstate of the corresponding cell, which is formed by these two opposing faces~\cite{kubica2018abcs}. If all three pairs of opposing faces agree, we know that either no data error or an error on the corner qubit of the yellow cell (qubit $8$) has occurred and we can proceed. If we find some disagreement, we know that an error has happened on a specific pair of qubits. To localize the error, we measure the yellow X-cell. If we measure $-1$, we know that an the error has occurred on the corresponding qubit of the yellow cell and can correct for it and continue. If we find $+1$, we know that the error has occurred on the Steane code and update the syndrome accordingly. In practice, the bulk, which is formed by those data qubits of the tetrahedron that are not part of the Steane code (yellow cell), can be measured destructively to localize all potentially dangerous errors on the data qubits.

For switching back to the tetrahedral code afterwards, the bulk has to be in a specific state that corresponds to the $+1$-eigenstate of its cells and faces. 
To this end, the bulk has to be re-initialized into the correct state using the circuit given in Fig.~\ref{fig:encoding_bulk}. For switching from the Steane code to the $[[15, 1, 3]]$ code, we identify only three qubit positions on which a single X-error can lead to a logical failure indicated in Fig.~\ref{fig:dangerous_errors_switching}(b), by numerically checking all possible error positions. We can distinguish these X-errors by measuring the complete Z-syndrome of the Steane code. To avoid further error propagation, we use the flag qubit scheme for this syndrome measurement. If a circuit flags, we measure one of the weight-eight X-cells using a single ancilla qubit to localize dangerous flag errors. After identifying the dangerous flag error, we remeasure the syndrome once without flags, to ensure that the syndrome contains the detected error. We can then update the syndrome according to the detected error correctly and continue. 

\subsection{Measurement errors}\label{sec:errors_measurement}

A single error on the measurement of the first ancilla qubit in the circuits in Fig.~\ref{fig:flag_measurements_circuits} would invert the measurement outcome and can lead to a logical failure after applying the corresponding switching operation. To achieve fault tolerance, we repeat the syndrome measurement and take a majority vote. If we obtain the same syndrome twice, we assume that no single measurement error has occurred and continue. If the syndromes are different, we repeat the syndrome measurement a third time, without a flag qubit, and proceed with the result of this last measurement. The probability that the measurement is incorrect two times is order $p^2$, where $p$ is the failure probability for a single measurement~\cite{Nielsen_and_Chuang}.

\section{Non-deterministic FT code switching}\label{sec:morphing}

The above deterministic schemes for code switching require a large number of gates, as summarized in Tab.~\ref{tab:resources_FT_blocks}, and a large circuit depth due the additional checks that have to be performed. 
We propose non-deterministic schemes as more feasible alternatives on near-term devices. These schemes achieve lower failure rates than the deterministic protocols, due to effectively postselecting for errors and reducing the required number of CNOT-gates and, therefore, the circuit depth. However, postselecting results in a finite success rate, since a fraction of runs is discarded. 
These non-deterministic schemes allow, for example, the preparation of a magic state on a two-dimensional color code with fewer resources than for the deterministic scheme while achieving lower logical failure rates. 
The methods we describe in the following are postselection and code morphing to an effective error detecting code. 

\subsection{Code switching with postselection}\label{sec:postselection}

For FT switching between the Steane and the $[[15, 1, 3]]$ code, we check for potentially dangerous errors on data and ancilla qubits. As discussed in Sec.~\ref{sec:FT_CS}, this is done by measuring extra stabilizers in addition to the gauge operators and using flag qubits. We now adjust the scheme to postselect for any detected error: whenever an error is detected, we now stop the protocol and discard the results. This includes flag qubits measured in the $\ket{1}$- or $\ket{-}$-state and any disagreement in opposing faces for the data error check. By postselecting~\cite{aharonov1988result, peres1989quantum} for any detected error, a fraction of weight-$2$ or higher weight errors is sorted out, resulting in lower logical failure rates. The number of required qubits and two-qubit gates remains the same as for the fault-free deterministic protocols, since the exact same circuits are executed.

\subsection{Code switching with morphed codes}

\textit{Code morphing} was introduced in~\cite{vasmer2022morphing} and is a method for generating new codes from existing ones by effectively turning the logical qubits of a sub-code, called the \textit{child} code, into bare physical qubits. This new code inherits the logical gates of the \textit{parent} code. By morphing a suitable code, it is possible to find a code that still has a fault-tolerant T-gate, but requires fewer qubits and has lower stabilizer weights, while the code distance is reduced. Therefore, the number of required CNOT-gates for the additional stabilizer measurements for error-checks can be reduced. 
This offers a more feasible alternative for code switching on current experimental setups. 

\subsection*{Morphing the tetrahedral code $[[15, 1, 3]]$}
The $[[15, 1, 3]]$ code contains a smaller stabilizer code on, for example, the yellow cell of the code. This smaller sub-code  corresponds to a  $[[8, 3, 2]]$ code and is called the child code. This $[[8, 3, 2]]$ code has a transversal implementation of the CCZ gate~\cite{campbell2016smallest}. 

\begin{figure}[t]
     \centering
     \includegraphics[width=0.49\textwidth]{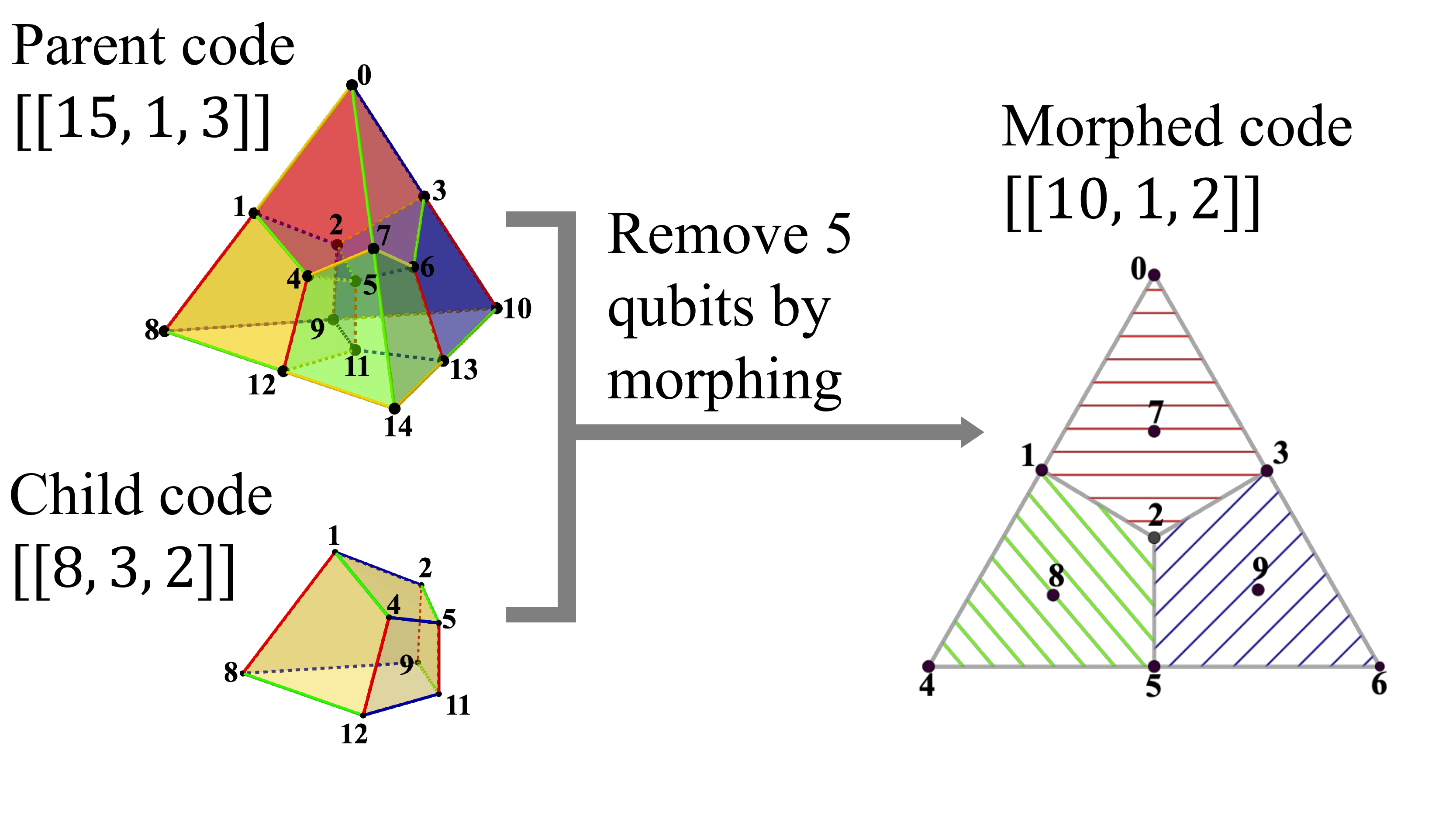}
    \caption{\justifying \textbf{Morphing the tetrahedral $[[15, 1, 3]]$ into the $[[10, 1, 2]]$ error detecting code. } One can morph the tetrahedral $[[15, 1, 3]]$ code with the yellow cell $[[8, 3, 2]]$, contained in the initial parent code. By taking the encoding circuit of the $[[8, 3, 2]]$ child code and applying its inverse to a codestate of the $[[15, 1, 3]]$ parent code, $5$ qubits are decoupled while keeping the encoded information in the remaining qubits. The encoding circuit and the inverted operation are discussed in App.~\ref{app:morphing}.
    This morphing effectively reduces the number of qubits and the stabilizer weights while reducing the distance. 
    Three physical qubits of the yellow cell remain in the final morphed code, which are depicted in the center of the plaquettes as qubits $7, 8$ and $9$. We obtain the $[[10, 1, 2]]$ code, which has new stabilizers of lower weight than the tetrahedral parent code. Instead of the weight-$8$ cells of the parent code, the cells of the morphed $[[10, 1, 2]]$ code now have weight 5 with $B'^X_R = X_0 X_1 X_2 X_3 X_7$, $B'^X_B = X_2 X_3 X_5 X_6 X_9$ and $B'^X_G = X_1 X_2 X_4 X_5 X_8$. The Z-stabilizers of the morphed code are defined on the three weight-$4$ plaquettes $Z_0 Z_1 Z_2 Z_3$, $Z_1 Z_2 Z_{4} Z_{5}$ and $Z_2 Z_3 Z_{5} Z_{6}$, as well as the three weight-$3$ operators $Z_2 Z_5 Z_7$ and $Z_1 Z_2 Z_9$ as well as $Z_2 Z_3 Z_8$. }
    \label{fig:morphing}
\end{figure}

One can invert the encoding circuit of the  $[[8, 3, 2]]$ code, by reverting the order of gates and the direction of all CNOT-gates, and apply it to the tetrahedral code. 
Thereby, the three encoded qubits of the $[[8, 3, 2]]$ are replaced by bare, physical qubits, leaving the new morphed $[[10, 1, 2]]$ code, as illustrated in Fig.~\ref{fig:morphing} and discussed in App.~\ref{app:morphing}. 
The new stabilizers of the morphed code can be derived by replacing the logical operators of the child code with the corresponding operators acting on the bare physical qubit. 
The previous weight-$8$ X-stabilizers of the parent code are turned into weight-$5$ cells on the morphed $[[10, 1, 2]]$ code. We illustrate the weight-$5$ operators by placing one extra qubit in the center of the plaquettes of the Steane code, as depicted in Fig.~\ref{fig:morphing} on the right. These are
\begin{align}
	B'^X_R &= X_0 X_1 X_2 X_3 X_7 \nonumber \\ B'^X_B &= X_2 X_3 X_5 X_6 X_9 \\ B'^X_G &= X_1 X_2 X_4 X_5 X_8 \nonumber 
\end{align}
Three Z-stabilizers are defined on the plaquettes of the code, and three Z-stabilizers are defined on the extended plaquette intersections as
\begin{align}
	B'^Z_R = Z_0 Z_1 Z_2 Z_3, \quad  B'^Z_{GB} = Z_2 Z_5 Z_7 \nonumber \\
	B'^Z_B = Z_2 Z_3 Z_5 Z_6, \quad  B'^Z_{RG} = Z_1 Z_2 Z_9 \\
	B'^Z_G = Z_1 Z_2 Z_4 Z_5, \quad  B'^Z_{RB} = Z_2 Z_3 Z_8. \nonumber 
\end{align}
By morphing a code, we can effectively trade off different code properties. In this case, we traded off the reduced stabilizer weights against code distance, since the morphed code is only an error detecting code with distance $d' = 2$. Note that code morphing is not physically executed but is a strategy to generate a new code that can be then used for code switching protocols.  

\begin{figure}[t]
     \centering
     \includegraphics[width=0.5\textwidth]{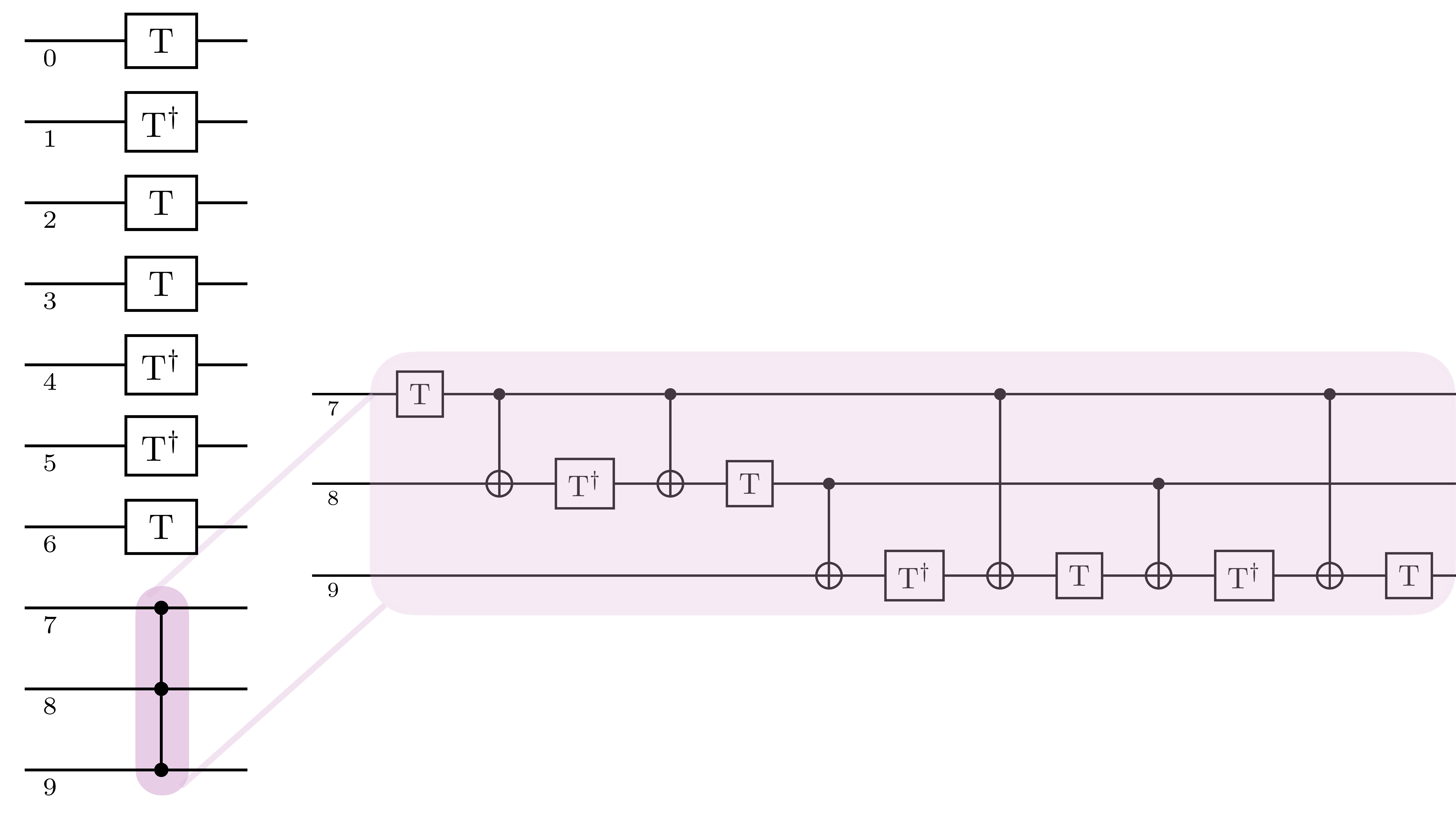}
     \caption{\justifying \textbf{Implementation of the FT logical $\overline{\mathrm{T}}$-gate for the $[[10, 1, 2]]$ code~\cite{vasmer2022morphing} }. Single T- and T$^{\dag}$-gates are applied to qubits $0 - 6$ and a CCZ operation is applied to the three qubits in the center of the plaquettes. The CCZ-gate on qubits $7, 8$ and $9$ can be decomposed into single-qubit and CNOT-gates~\cite{barnes2017fast} using the implementation given on the right. This implementation of $\overline{\mathrm{T}}$ with a three-qubit gate is still fault-tolerant even though errors on qubits $7, 8$ or $9$ can propagate onto other data qubits. Any Z-error on these qubits is detectable and all possible X-errors on these qubits are correctable. }
    \label{fig:T_gate_10_circuits}
\end{figure}

The logical $\overline{\mathrm{T}}$-gate can be implemented on the $[[10, 1, 2]]$ code using the circuit shown in Fig.~\ref{fig:T_gate_10_circuits}(a). This implementation is not transversal anymore but fault-tolerant in the sense that all possible errors resulting from a single Z-error can be detected. All possible X-errors on qubits $7, 8$ and $9$, as for example $X_7 X_{8} X_9$, are correctable because the corresponding Z-syndromes do not coincide with any single-error configuration. Therefore, we can assign the syndromes to these higher weight X-errors and correct for them. This implementation of $\overline{\mathrm{T}}$ can be understood by interpreting code morphing as a replacement of the logical qubits, encoded in the yellow cell, by three bare physical qubits. The action of the T- and T$^{\dag}$-gates in the FT implementation of the logical $\overline{\mathrm{T}}$-gate on the yellow cell corresponds to a logical $\overline{\mathrm{CCZ}}$ on the three logical qubits of the child code $[[8, 3, 2]]$~\cite{campbell2016smallest}. By morphing the tetrahedral code, this logical operation $\overline{\mathrm{CCZ}}$ is replaced by the corresponding action on the three bare physical qubits, which is the CCZ-gate on $7, 8$ and $9$. The CCZ-gate itself can be implemented using only single T-, T$^{\dag}$- and CNOT-gates as shown in Fig.~\ref{fig:T_gate_10_circuits}(b). The morphed $[[10, 1, 2]]$ code is the smallest known stabilizer code that has a FT \mbox{T-gate}.

\subsection*{FT code switching between $[[10, 1, 2]]$ and $[[7, 1, 3]]$}

In order to switch between the two stabilizer codes, first, the two codes must have a common representation of the logical operators and, secondly, a subsystem code with stabilizer group $\mathcal{S}$ has to be defined that suffices $\mathcal{S} \subset \mathcal{S_{\mathrm{[[10, 1, 2]]}}}$ and $\mathcal{S} \subset (\mathcal{S}_{\mathrm{Steane}}, \mathcal{S}_{\mathrm{bulk}})$. The $[[10, 1, 2]]$ code inherits the logical operators from its parent code, the $15$-qubit tetrahedral code. Since the $15$-qubit tetrahedral code and the Steane code have a common representation of their logical operators, this condition is consequentially also fulfilled for the $[[10, 1, 2]]$ code. 
Furthermore, a suitable subsystem can be defined whose gauge group $\mathcal{G}$ is generated by the stabilizers of the $[[10, 1, 2]]$ code, which includes the Z-plaquettes of the Steane code, and the single-qubit Pauli operators $X_7, X_8$ and $X_9$ for the indexing given in Fig.~\ref{fig:morphing}. The stabilizer group $\mathcal{S}$ of the subsystem is then generated by the weight-$5$ X- and Z-cells of the $[[10, 1, 2]]$ code structure. 

We can switch from the $[[10, 1, 2]]$ code to the Steane code by measuring the stabilizers of the Steane code, which are not fulfilled by the initial codestate in the $[[10, 1, 2]]$ code. These stabilizers are the X-plaquettes of the Steane code, since the $[[10, 1, 2]]$ code only fulfills the weight-$5$ X-cells and not the weight-$4$ X-plaquettes. After measuring the stabilizers of the target code, we apply combinations of the gauge operators $B'^Z_{BG} = Z_2 Z_5 Z_7$, $B'^Z_{RG} =  Z_1 Z_2 Z_9$ and $B'^Z_{RB} =  Z_2 Z_3 Z_8$, in order to fix the gauge state accordingly. 

The FT switching scheme from the $15$-qubit tetrahedral code to the Steane code, discussed in the previous section, can directly be adapted for the morphed code. The X-plaquettes of the Steane code are measured using flag qubits and errors on data qubits can, again, be detected by measuring opposing pairs of stabilizers in the subsystem. We only need to replace the bulk measurement with single-qubit measurements of qubits $7, 8$ and $9$ in the X-basis. If some disagreement between opposing pairs of stabilizers, as for example between $X_7$ and $B'^X_R = X_0 X_1 X_2 X_3$, is found, we have to discard the corresponding run because Z-errors on the data qubits cannot be uniquely identified in the subsystem which makes the protocol non-deterministic. 

For switching from the Steane code to the $[[10, 1, 2]]$ code, we have to measure those stabilizers of the $[[10, 1, 2]]$ code, which are not fulfilled in the Steane code. These are the weight-$3$ Z-stabilizers on the extended intersections of plaquettes $(B'^Z_{RB}, B'^Z_{GB}, B'^Z_{RG})$. Based on the measurement outcome, we apply combinations of the Steane X-plaquettes. Since the stabilizers are of weight-$3$, any error on the ancilla qubit during these measurements may indeed propagate but is only equivalent to a single error on the data qubits. So, for a FT measurement of the weight-$3$ stabilizers, we do not need additional flag qubits. To check for errors on the data qubits that can lead to a logical failure, we measure the green and blue Z-plaquettes. If some error is discovered in this check, we discard this run. If no error is found, we take the measured Z-syndrome and continue. These measurements can also be performed without additional flag qubits, since every dangerous flag error is detectable on the target code. 

In the following section, we present simulation results for code switching protocols using the tetrahedral $[[15, 1, 3]]$ and the morphed $[[10, 1, 2]]$ code.

\section{Error model and simulation methods}

In the following simulations, we investigate the logical failure rates $p_L$ for different protocols by performing Monte Carlo (MC) simulations. We consider circuit-level noise with a single error parameter $p$, which describes the error rates on each component. This includes single- and two-qubit gates, as well as measurements and initializations of physical qubits. Each circuit element is modeled as an ideal operation $U_{\mathrm{ideal}}$ followed by an error $E$ with the given probability
\begin{align}
    U_{\mathrm{faulty}} = E\cdot U_{\mathrm{ideal}}. 
\end{align}
Furthermore, faulty operators for measurements are placed before the ideal measurement location. 
We consider depolarizing noise channels on all single- and two-qubit gates, which are described by the noise operators~\cite{bermudez2017assessing, postler2022demonstration, parrado2021crosstalk}
\begin{align}
	E_1 &\in \{ \sigma_k, \forall k \in \{1, 2, 3 \} \} \\
	E_2 &\in \{\sigma_k \otimes \sigma_l, \forall k, l \in \{0, 1, 2, 3 \}  \} \backslash  \{I \otimes I \}
\end{align}
with the Pauli matrices $\sigma_k = \{I, X, Y, Z \}$ with $k=0, 1, 2, 3$. 
The depolarizing channels are then determined by
\begin{align}
    \epsilon_1(\rho) &= (1 - p)\rho + \frac{p}{3} \sum_{i= 1}^3 E^{i}_1 \rho E^{i}_1 \\
    \epsilon_2(\rho) &= (1 - p)\rho + \frac{p}{15} \sum_{i= 1}^{15}   E_2^{i} \, \rho\, E_2^{i}.  \label{eq:depol_single_qubit}
\end{align}

This means that, one of the $3 (15)$ possible combinations of single(two)-qubit Pauli-errors is applied with a given probability $p/3$ ($p/15$). Qubits are prepared in $\ket{0}$ and measured in the Z-basis. We model faults on these state preparations and measurements by applying X-errors after state preparations and before measurements each with a probability $p$. 

The logical failure rate is determined as follows. At the end of each MC shot, we perform one round of ideal QEC on the noisy output states, which maps the state back into the codespace. Finally, we extract the expectation value of the corresponding logical operator $\langle \overline{O}\rangle $ classically in software. This indicates that a logical error has occurred if we find an expectation value of $-1$. We realize $n_{\mathrm{runs}}$ MC shots for each protocol and obtain the logical failure rate $p_L$ by averaging over all shots. 

The uncertainties on the logical failure rates are calculated by taking the uncertainty of the mean value for a binomial distribution 
\begin{align}
	\sigma_{p_L} = \sqrt{\frac{p_L(1 - p_L) }{n_{\mathrm{runs}}}}
\end{align}
where $n_{\mathrm{runs}}$ is the total number of simulation runs and $p_L$ is the estimated logical failure probability. We repeat each simulation $10^5$ to $10^6$ times, until the relative uncertainty on a given data point is smaller than $5\%$. Up to $10^6$ shots are required for small physical error rates, where error events are rare. 

We use the package \textsc{Pecos}, which is a Python framework for studying, developing, and evaluating quantum error-correction protocols through numerically performing stabilizer or statevector simulations of noisy quantum circuits~\cite{pecos_git}. 

\section{Building blocks using FT code switching}\label{sec:results}

\begin{figure*}[t]
     \centering
     \includegraphics[width=\textwidth]{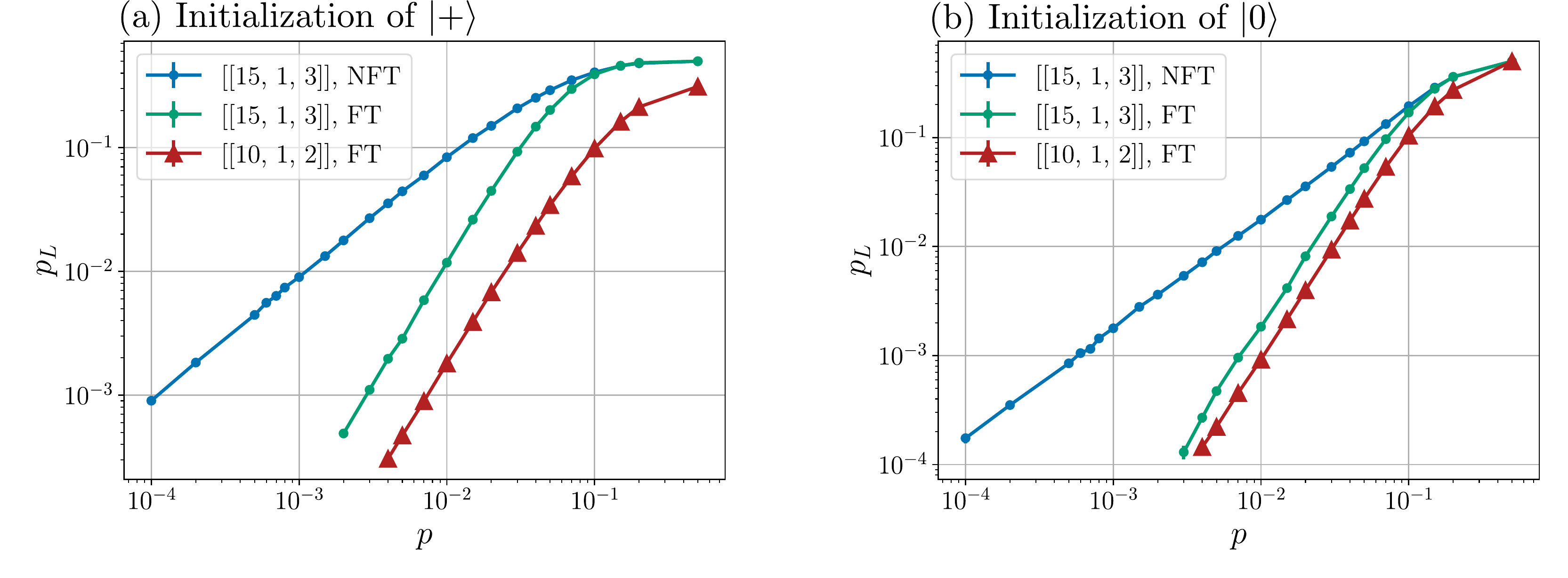}
    \caption{\justifying \textbf{Logical failure rates for the initialization of logical states illustrated in Fig.~\ref{fig:protocols_overview}(b).} The logical failure rates are shown for the initialization of (a) $\ket{\overline{0}}$ and (b) $\ket{\overline{+}}$ on the tetrahedral code $[[15, 1, 3]]$ using a FT scheme that includes a verification step (green) and a Non-FT (NFT) scheme without this verification (blue), as well as on the morphed $[[10, 1, 2]]$ code (dark red). 
    The logical failure rates for the Non-FT protocol scale linearly in the error parameter $p$, since single physical errors can result in a logical failure. For the FT initializations, we can identify a quadratic scaling in $p$, verifying that only two physical failures introduce a logical error. For the $[[10, 1, 2]]$ code, lower failure rates are reached than for the initialization of the same state on the tetrahedral code. }
    \label{fig:init}
\end{figure*}

\begin{figure*}[t]
     \centering
	\includegraphics[width=\textwidth]{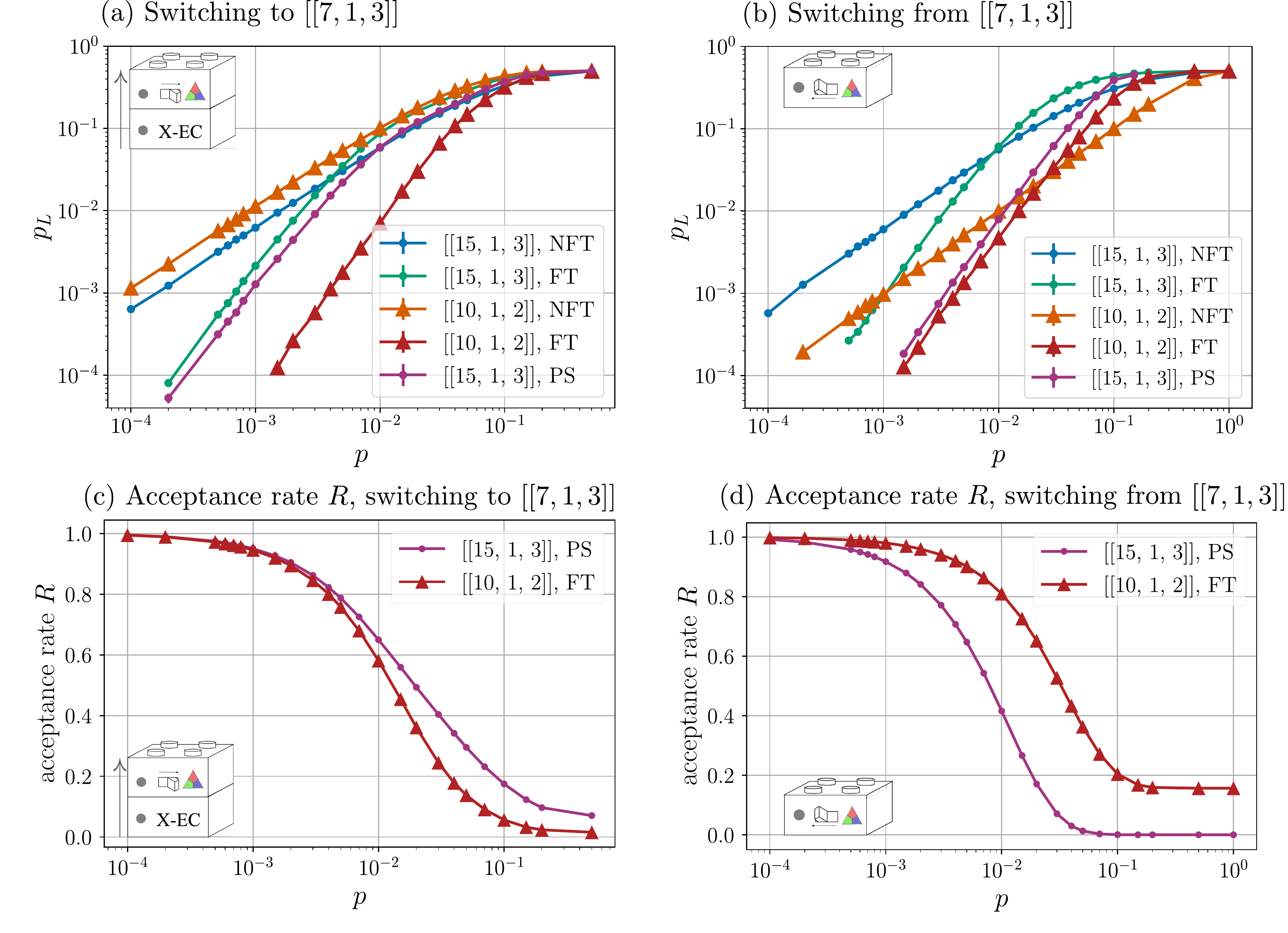}
    \caption{\justifying \textbf{Logical failure rates and acceptance rates $R$ for building blocks introduced in Fig.~\ref{fig:protocols_overview}(b)}. (a) Logical failure rates for switching from the Steane code to the tetrahedral code using a fault-tolerant (FT, green) scheme, a Non-FT (NFT, blue) scheme and postselection (PS, purple), as well as for switching from the Steane code to the morphed code using a FT (dark red) and NFT (orange) scheme. The FT non-deterministic protocols achieve lower logical failure rates than the deterministic ones, while the FT scheme with the $[[10, 1, 2]]$ code performs best. The circuit depth of the required round of X-error correction on the tetrahedral code before switching to the Steane code is large, as summarized in Tab.~\ref{tab:resources_FT_blocks}. For switching from the $[[10, 1, 2]]$ code, the circuit depth of one round of X-error correction is a factor of $3$ smaller, which reduces the number of potential error locations and, therefore, lowers the logical failure rate. (b) Logical failure rates for switching from the tetrahedral code to the Steane code using a FT (green) scheme, a Non-FT (blue) scheme and postselection (purple), as well as for switching from the morphed code to the Steane code using a FT (dark red) and Non-FT (orange) scheme. Switching from the $[[10, 1, 2]]$ code achieves lower failure rates for $p>10^{-3}$, even for the Non-FT scheme. (c) Acceptance rates $R$ for switching from the Steane code to the $[[15, 1, 3]]$ code and the $[[10, 1, 2]]$ code using non-deterministic protocols. The acceptance rates for the $[[10, 1, 2]]$ code are slightly smaller than for the $[[15, 1, 3]]$ code in the considered range of the physical error rate $p$. (d) Acceptance rates $R$ for the inverse switching direction. The acceptance rates for the $[[10, 1, 2]]$ code are higher than for the $[[15, 1, 3]]$ code in the considered range of the physical error rate $p$. 
    }
    \label{fig:switching}
\end{figure*}

\begin{figure*}[t]
     \centering
	\includegraphics[width=\textwidth]{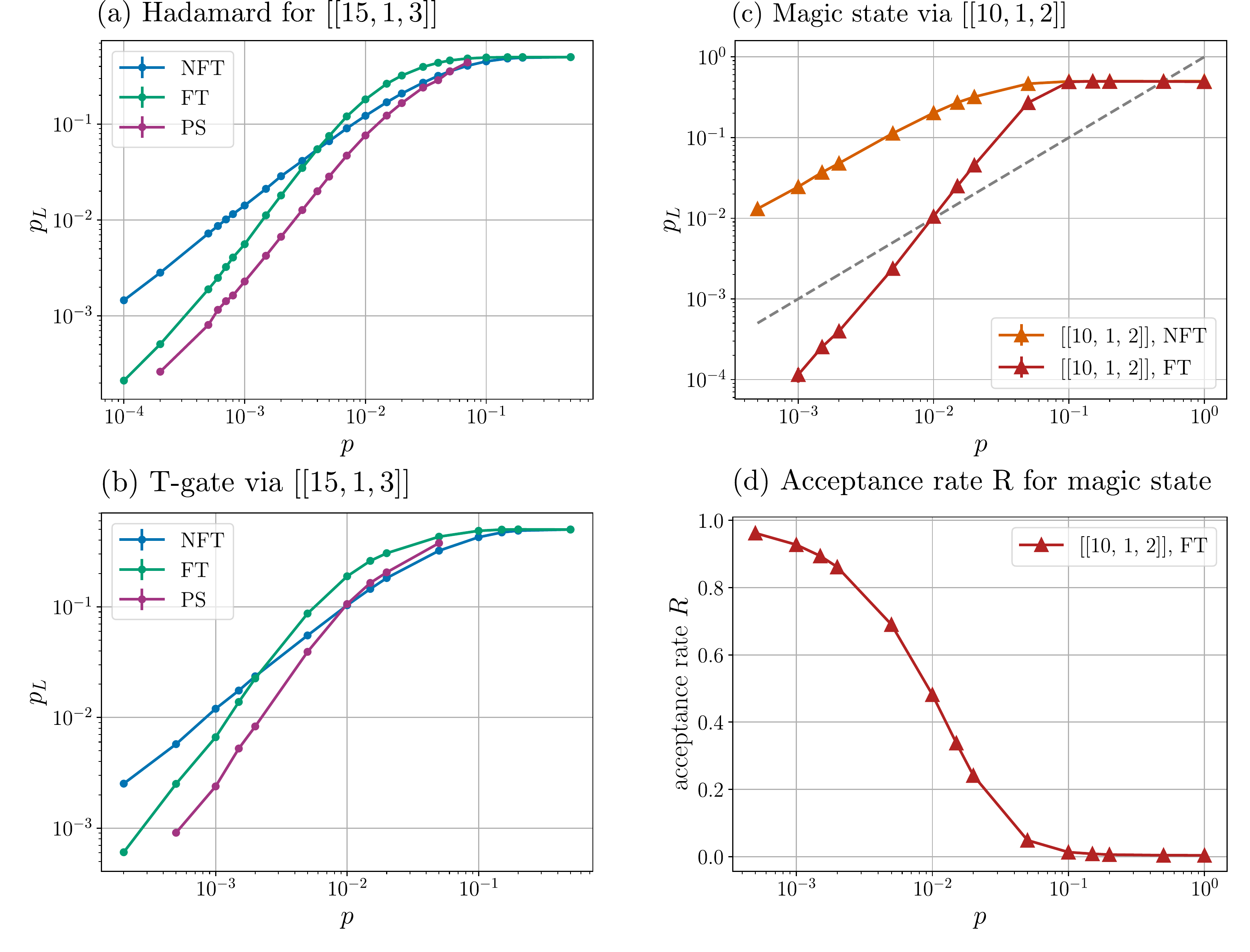}
    \caption{\justifying \textbf{Logical failure rates for logical operations illustrated in Fig.~\ref{fig:protocols_overview}(c).} The logical failure rates are averaged over different initial states for the Non-FT (blue), FT (green) and postselected (PS, purple)  implementation of (a) the Hadamard-gate on the tetrahedral $[[15, 1, 3]]$ code and (b) the T-gate on the Steane code. The Non-FT implementations of both gates achieve similar logical failure rates, while the FT versions perform worse for the T-gate than for the Hadamard-gate. Note that for a specific input state, these failure rates may look different. Different states are sensitive to different kinds of errors and the amount of dangerous error positions throughout the protocols varies for each type of error. (c) Logical failure rates for the preparation of the magic state $\overline{T}|\overline{+}\rangle$ on the $[[7, 1, 3]]$ Steane code using the morphed $[[10, 1, 2]]$ code for the Non-FT (orange) and FT (red) scheme. The grey dotted line corresponds to the physical error rate $p$. For $p < 10^{-2}$, the FT scheme the logical failure rate is smaller than $p$. (d) Acceptance rates $R$ for magic state preparation via the morphed $[[10, 1, 2]]$ code. For $p < 10^{-2}$, more than half of the runs is accepted. }
    \label{fig:protocols_results}
\end{figure*}

We determine the logical failure rates for each block given in Fig.~\ref{fig:protocols_overview} by means of the above described simulation methods. This includes the initialization of the logical states $|\overline{0}\rangle$ and $|\overline{+}\rangle$ on the tetrahedral $[[15, 1, 3]]$ and the morphed $[[10, 1, 2]]$ code, FT switching in both directions to and from the $[[7, 1, 3]]$ Steane code, as well as the composite protocols specified in Fig.~\ref{fig:protocols_overview}(c). These protocols implement the Hadamard-gate on the tetrahedral code, the T-gate on the Steane code and the preparation of a magic state on the Steane code via the morphed code. 

For the initialization of logical states on the tetrahedral code $[[15, 1, 3]]$, we construct the FT circuits shown in Figs.~\ref{fig:init_tetra_logical_0_circuit} and \ref{fig:init_tetra_logical_plus_circuit}. For the FT initialization of a logical state on the morphed $[[10, 1, 2]]$ code, we design the circuits shown in Fig.~\ref{fig:10_code_circuits}. The logical failure rates for these protocols are shown in Fig.~\ref{fig:init}. In the regime of low physical error rates $p \rightarrow 0$, we identify a quadratic scaling in the logical failure rate $p_L \sim p^2$ for the FT protocols and a linear scaling $p_L \sim p$ for the Non-FT protocol. This indicates that for the Non-FT protocol, single errors result in a logical failure, whereas for the FT protocols only weight-$2$ error configurations contribute to $p_L$. 
The logical failure rates for the initialization of the morphed $[[10, 1, 2]]$ code is smaller than for the tetrahedral code. Since the circuit depth and the number of required gates for the $[[10, 1, 2]]$ code is smaller, as summarized in Tab.~\ref{tab:resources_FT_blocks}, there are fewer weight-$2$ error configurations that can contribute to the logical failure rate. Thus, the logical failure rates are lower than for the tetrahedral code, as can be seen in Fig.~\ref{fig:init}. 
    
The figures \ref{fig:switching}a and b show the logical failure rates for FT switching with the tetrahedral code. Furthermore, we determine the logical failure rates for Non-FT code switching, where the corresponding three stabilizers are measured once without flags. A third simulation considers switching with the tetrahedral code while postselecting for any detected error. Whenever a flag is triggered, or an error on a data qubit is detected, the corresponding run is discarded and the protocol is restarted as discussed in Sec.~\ref{sec:postselection}. 
For the morphed code, we determine the logical failure rates for FT switching, as well as for the Non-FT switching scheme. These protocols are non-deterministic, since the $[[10, 1, 2]]$ code is an error detecting code. For Non-FT switching, the stabilizers are measured once without flags.  

The logical failure rates for FT switching from the tetrahedral $[[15, 1, 3]]$ code to the $[[7, 1, 3]]$ Steane code intersects with the logical failure rate for Non-FT switching for values at approximately $p = 5\cdot 10^{-3}$. This crossing point is higher for the inverse direction at approximately $p = 10^{-2}$. 
The X-error correction block for switching to the Steane code, which is necessary to achieve fault tolerance, requires many CNOT gates, as summarized in Tab.\ref{tab:resources_FT_blocks}. This increases the number of possible error configurations and, therefore, leads to an asymmetry between the two switching directions. 
Considering switching with the tetrahedral $[[15, 1, 3]]$ code, we observe that the Non-FT schemes achieve similar values for $p_L$ for both switching directions, while for the FT protocols, these rates differ. The Non-FT protocol is symmetric in the switching directions. In both cases, the three weight-$4$ stabilizers have to be measured and each Pauli error is equally likely in this error model. For the FT switching protocols, the directions are not symmetric. 
For Non-FT switching with the morphed $[[10, 1, 2]]$ code, the logical failure rates also differ. Non-FT switching from the $[[10, 1, 2]]$ code to the Steane code achieves higher logical failure rates than the inverse direction, since the stabilizer weights for the two directions differ. For switching to the Steane code, the weight-$4$ plaquettes are measured, where for example a single error on an ancilla qubit can cause a logical failure. For switching from the Steane code to the $[[10, 1, 2]]$ code, only the weight-$3$ stabilizers have to be measured. 

The figures \ref{fig:switching}(c) and (d) show the acceptance rates for the non-deterministic protocols, which include switching with the tetrahedral code with postselection and FT switching with the morphed code. The acceptance rate 
\begin{align}
	R = \frac{n_{\mathrm{runs, accepted}}}{n_{\mathrm{runs, total}}}
\end{align}
indicates the number of runs that are not discarded due to an error-detection event divided by the total number of runs. 
For switching from the $[[7, 1, 3]]$ Steane code to the morphed $[[10, 1, 2]]$ code, a larger fraction of runs is accepted than for the corresponding protocol with the tetrahedral $[[15, 1, 3]]$ code. This is inverted for the reverse direction, while the relative difference between the acceptance rates decreases. 

Figures \ref{fig:protocols_results}(a) and (b) show the logical failure and acceptance rates for the gates that cannot be realized transversally in the corresponding codes but are necessary in order to complete the universal gate set. For the Hadamard-gate on the tetrahedral $[[15, 1, 3]]$ code, first, switching to the $[[7, 1, 3]]$ Steane code is applied, followed by a transversal Hadamard operation on the Steane code and switching back to the tetrahedral code. Analogously, the T-gate on the Steane code is implemented by switching to the tetrahedral code, applying the transversal T-gate and switching back to the initial code. Again, we consider FT protocols, as discussed in the previous sections, as well as Non-FT switching and a non-deterministic version of the corresponding protocol, which includes postselection for any detected error. 

\begin{figure*}[t]
     \centering
      \includegraphics[width=\textwidth]{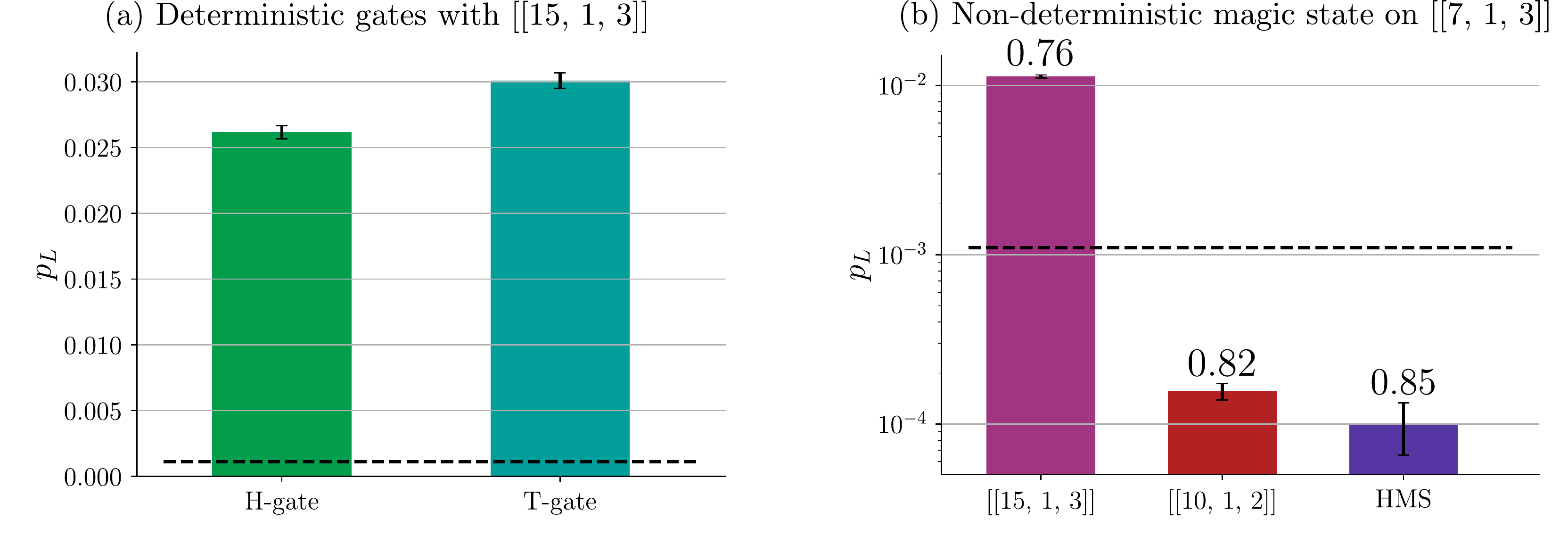}
     \caption{\justifying \textbf{Projected performance of logical operations on trapped ion quantum processors.} (a) The projected performance for deterministic Hadamard- (left) and T-gate (right) is estimated for current ion trap setups~\cite{andersen2020repeated}. These protocols correspond to the two left composite column in Fig.~\ref{fig:protocols_overview}(c). The shown logical failure rates are calculated for an ideal input state. 
     The left bar corresponds to the averaged logical failure rate of $p_L = 2.62\cdot 10^{-2} \pm 0.05 \cdot 10^{-2}$ for a Hadamard-gate on the $[[15, 1, 3]]$ tetrahedral code via the Steane code. For the T-gate on the Steane code, we obtain slightly higher failure rates of $p_L = 3.01 \cdot 10^{-2}\pm 0.06\cdot 10^{-2}$. The black dashed line corresponds to the failure rate of the according sequence for a single physical qubit. 
     For both logical gates, the determined logical failure rate on near-term ion trap processors is still more than one order of magnitude larger than the expected rate for a single physical qubit. 
     (b) Logical failure rates for the preparation of the magic state $\overline{T}|\overline{+}\rangle$ on the Steane code on a logarithmic scale. These protocols correspond to the composite column in Fig.~\ref{fig:protocols_overview}(c) on the right, including a noisy initialization, a noisy application of the T-gate and noisy switching to the Steane code. For the tetrahedral $[[15, 1, 3]]$ code (left) with postselection, we obtain $p_L = 1.13 \cdot 10^{-2} \pm 0.02 \cdot 10^{-2}$ at an acceptance rate of $R = 76\%$. Using the morphed $[[10, 1, 2]]$ code (center) yields $p_L = 1.6\cdot 10^{-4} \pm 2\cdot 10^{-5}$ with an acceptance rate of $R = 82\%$. \textit{HMS} corresponds to the preparation of a heralded magic state on the Steane code using a state-of-the-art FT implementation~\cite{postler2022demonstration}. For this scheme, we obtain $p_L = 1.0\cdot 10^{-4} \pm 3\cdot 10^{-5}$ with an acceptance rate of $R = 85\%$. 
     }
    \label{fig:projected_performance}
\end{figure*}

The FT protocols only outperform the Non-FT schemes for $p \leq 5\cdot 10^{-3}$, due to the large overhead that is required to achieve fault tolerance. Again, we observe that the failure rates for the Non-FT scheme on the tetrahedral code are similar for both operations. In this case, the switching cycles are symmetric, meaning that after the first switching step in one protocol, the probability for a given number of errors is the same as after the first switching step for the other protocol. For the FT protocols, these may differ, since the two switching directions include different operations and a different number of stabilizer measurements. Non-FT switching from the Steane code to the morphed $[[10, 1, 2]]$ code achieves lower logical failure rates than for the inverse direction. For switching from the Steane code, we only need to measure the weight-$3$ Z-stabilizers. Any error on the ancilla qubits may propagate but is always equivalent to a weight-$1$ error. For switching in the inverse direction, the weight-$4$ plaquettes have to be measured requiring more two-qubit gates and giving rise to logical failures due to propagated errors on the ancilla qubits.

Figure \ref{fig:protocols_results}(c) and (d) show the logical failure rates and acceptance rates for the preparation of a magic state on the Steane code using the morphed $[[10, 1, 2]]$ code. The logical failure rates scale quadratically, indicating that no single error results in a logical failure. The logical failure rate of the FT scheme surpasses the corresponding physical error rate at approximately $p = 10^{-2}$, while keeping more than $50$\% of the runs below this point. 

To summarize, we observe the characteristic quadratic scaling at low physical error rates  for the constructed FT code switching protocols. In addition, we find that the Non-FT code switching protocol for switching from the $[[7, 1, 3]]$ Steane code to the morphed $[[10, 1, 2]]$ code achieves lower physical error rates than FT switching with the tetrahedral $[[15, 1, 3]]$ code for $p \leq 10^{-3}$. Furthermore, we estimate a breakeven point at a physical error rate of approximately $1 \cdot 10^{-2}$ with an acceptance rate of $50\%$ for the preparation of a magic state on the Steane code using the morphed $[[10, 1, 2]]$ code.

\section{Projected performance for trapped-ion quantum processors}\label{sec:projected_performance}

\begin{figure*}[t]
     \centering
      \includegraphics[width=\textwidth]{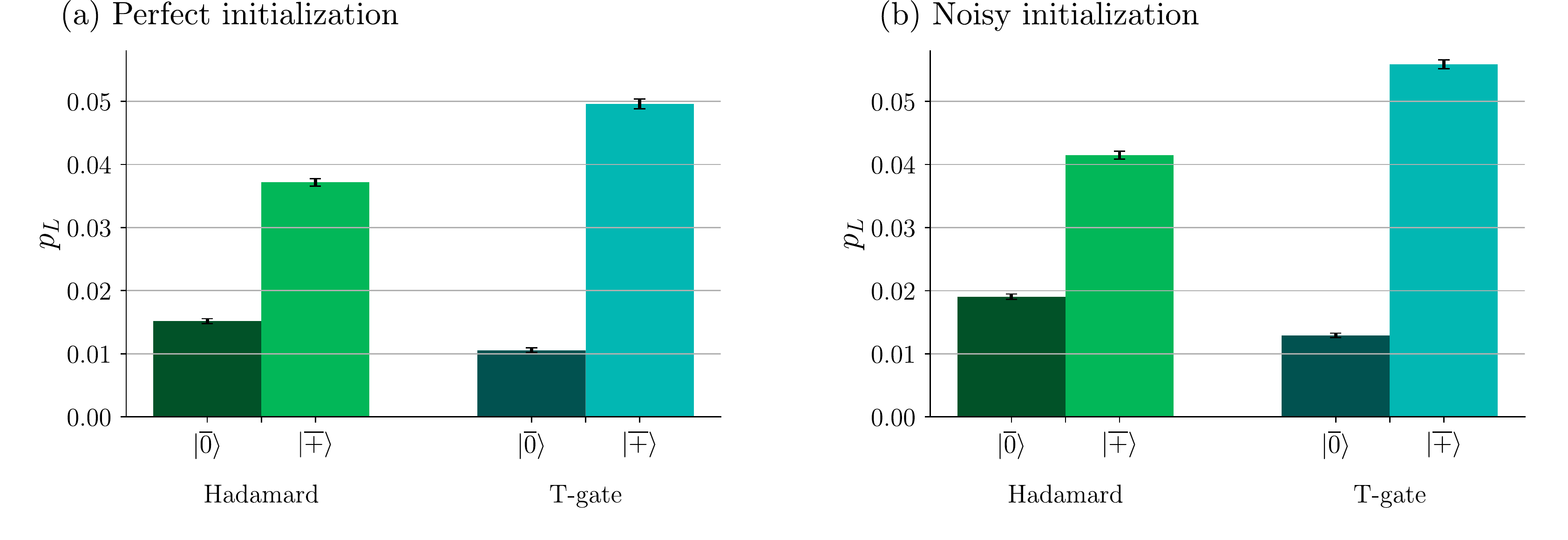}
     \caption{\justifying \textbf{Projected performance of deterministic logical gates for different encoded states.} Projected performance for deterministic Hadamard- (left) and T-gate (right) for current ion trap setups~\cite{andersen2020repeated} for logical input states $|\overline{0}\rangle$ and $|\overline{+}\rangle$ (a) for perfect and (b) noisy initialization. The Hadamard on a perfect state $|\overline{0}\rangle$ achieves a logical failure rate of $p_L = 1.52 \cdot 10^{-2} \pm 0.04 \cdot 10^{-2}$, while for $|\overline{+}\rangle$ we obtain a higher logical failure rate of $p_L = 3.72 \cdot 10^{-2} \pm 0.06 \cdot 10^{-2}$. Similarly for the T-gate, we find higher logical failure rates for $|\overline{+}\rangle$ with $p_L = 4.96 \cdot 10^{-2} \pm 0.08 \cdot 10^{-2}$ than for $|\overline{0}\rangle$ with $p_L = 1.05 \cdot 10^{-2} \pm 0.04 \cdot 10^{-2}$. The noisy initialization is performed using the circuits given in Fig.~\ref{fig:init_tetra_logical_0_circuit}, \ref{fig:init_tetra_logical_plus_circuit}, \ref{fig:encoding_bulk}, \ref{fig:10_code_circuits} and \ref{fig:Steane_init}. For the noisy initialization, the Hadamard-gate on the logical state $|\overline{0}\rangle$ achieves a logical failure rate of $p_L = 1.90 \cdot 10^{-2} \pm 0.04 \cdot 10^{-2}$ and a rate of $p_L = 4.15 \cdot 10^{-2} \pm 0.06 \cdot 10^{-2}$ for $|\overline{+}\rangle$. For the T-gate on the Steane code, we find a failure rate of $p_L = 1.29 \cdot 10^{-2}\pm 0.04 \cdot 10^{-2}$ for $|\overline{0}\rangle$ and $p_L = 5.59 \cdot 10^{-2} \pm 0.07 \cdot 10^{-2}$ for $|\overline{+}\rangle$. }
    \label{fig:projected_performance_variants}
\end{figure*}

\begin{figure*}[t]
     \centering
      \includegraphics[width=\textwidth]{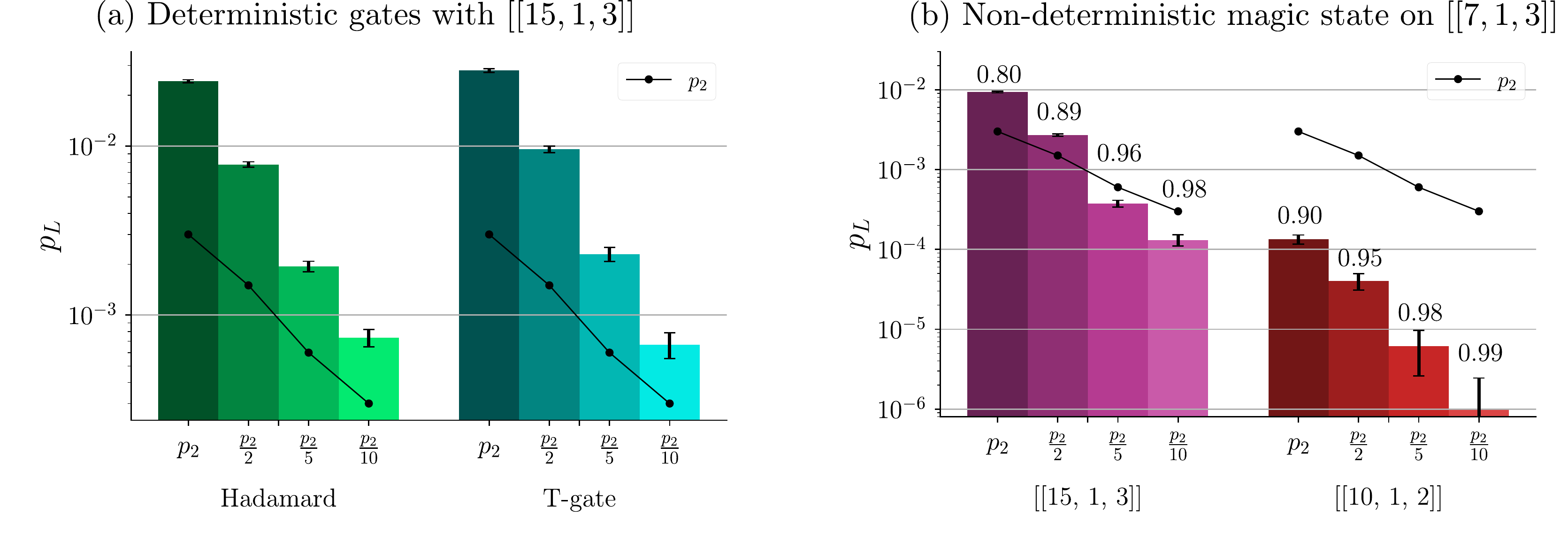}
     \caption{\justifying \textbf{Scaling of logical failure rates with the two-qubit gate noise parameter $p_2$ for different protocols.} (a) Averaged logical failure rates for the deterministic Hadamard-gate on the $[[15, 1, 3]]$ tetrahedral code and the deterministic T-gate on the Steane code according to the protocol specified in Fig.~\ref{fig:protocols_overview}(c) given an ideal logical input state. The parameters $p_1$, $p_i$ and $p_m$ are set to $0$ and $p_2$ is reduced by a factor $n$ for $n = 1, 2, 5, 10$. At the initial value of $p_2$, we find a logical failure rate of $p_L = 2.41 \cdot 10^{-2}\pm 0.05 \cdot 10^{-2}$ for the Hadamard-gate and $p_L = 2.79 \cdot 10^{-2} \pm 0.07 \cdot 10^{-2}$ for the T-gate. The black dots correspond to the physical error rate of $\frac{p_2}{n}$. The logical failure rates are larger than this physical error rate for the considered variation in $p_2$. 
      (b) Averaged logical failure rates for the preparation of a magic state on the Steane code on a logarithmic scale as specified in Fig.~\ref{fig:protocols_overview}(c). The preparation of the magic state includes the noisy initialization of the $[[15, 1, 3]]$ (left) and $[[10, 1, 2]]$ (right) code, the application of the FT T-gate and switching to the $[[7, 1, 3]]$ Steane code. We consider the same variation in the parameter $p_2$ for $p_1 = p_i = p_m = 0$. At the initial value of $p_2$, we find a logical failure rate of $p_L = 9.34 \cdot 10^{-3}\pm 0.20 \cdot 10^{-3}$ for the $[[15, 1, 3]]$ code and $p_L = 1.34 \cdot 10^{-4} \pm 0.17 \cdot 10^{-4}$ for the $[[10, 1, 2]]$ code. For magic state preparation using the tetrahedral $[[15, 1, 3]]$ code and postselection, the logical failure rates succeeds the corresponding physical failure rate at $\frac{p_2}{5}$, while for the morphed $[[10, 1, 2]]$ code, $p_L$ is already below this rate for the initial value of $p_2$. }
    \label{fig:p2_sensitivity}
\end{figure*}

Noise on real devices cannot be described by a single-parameter noise model. Error rates can vary for different circuit components, for example the two-qubit gate error rates are typically higher than those of other error sources. In order to estimate the projected performance on near-term trapped-ion quantum processors, we consider a multi-parameter noise model with different errors rates for each type of circuit component~\cite{postler2022demonstration, andersen2020repeated, trout2018simulating, heussen2023strategies}. 

We simulate a Hadamard-gate for the tetrahedral code, a T-gate for the Steane code and the preparation of a magic state on the Steane code using a modified noise model inspired by current ion trap processors. Instead of using a single error parameter $p$, we use multiple parameters to describe noise on different components of the circuit. We specify the depolarizing channel on single-qubit gates with a parameter $p_1$ and on two-qubit gates with $p_2$. Faulty measurements and initializations are modeled with faults with a probability $p_m$ and $p_i$, respectively, with which the state of the qubit is inverted. In the following simulations, we choose parameter values based on recent benchmarks on ion trap processors~\cite{ryan2021realization} as $p_1 = 1\cdot 10^{-4}, p_2 = 3\cdot 10^{-3}, p_i = 1\cdot 10^{-3}$ and $p_m = 1\cdot 10^{-3}$. A detailed discussion on the modeling of noise in ion traps~\cite{parrado2021crosstalk, ryan2021realization, bermudez2017assessing} is beyond the scope of this work. x

Fig.~\ref{fig:projected_performance}(a) shows the averaged logical failure rates for deterministic gates using code switching protocols. We find a logical failure rate of $2.62 \cdot 10^{-2} \pm 0.05 \cdot 10^{-2}$ for the deterministic Hadamard-gate on the tetrahedral $[[15, 1, 3]]$ code and a failure rate of $3.01 \cdot 10^{-2}\pm 0.06 \cdot 10^{-2}$ for the deterministic T-gate on the Steane code for this noise model. Both these values are much larger than the estimated corresponding physical error rate, but are comparable to recent benchmarks on trapped-ion setups~\cite{postler2022demonstration, ryan2021realization, chamberland2019fault}. 

Fig.~\ref{fig:projected_performance}(b) shows the averaged logical failure rates for magic state preparation on the Steane code for three different protocols. The left bar corresponds to the initialization of $|\overline{+}\rangle$ on the $[[15, 1, 3]]$ tetrahedral code, the application of a T-gate and, then, switching to the Steane code while postselecting for errors on data and ancilla qubits. For this protocol, we find a logical failure rate of $1.13 \cdot 10^{-2}\pm 0.02 \cdot 10^{-2}$ for the specified noise parameters. Using the morphed $[[10, 1, 2]]$ code for the same protocol yields $p_L = 1.6\cdot 10^{-2} \pm 2 \cdot 10^{-2} \cdot 10^{-3} \cdot 10^{-2}$, which is below the corresponding failure probability of a single physical qubit. Using the morphed $[[10, 1, 2]]$ code for the preparation of a magic state reduces $p_L$ by two orders of magnitude compared to the postselected implementation for the tetrahedral $[[15, 1, 3]]$ code. 

In order to compare our results to existing methods, we simulate the preparation of a magic state on the Steane code using a state-of-the-art method~\cite{chamberland2019fault, goto2016minimizing, postler2022demonstration}. Here, the magic state is prepared non-fault-tolerantly on the Steane code, followed by a measurement of a logical operator and one round of error detection on the Steane code. For this implementation, we find $p_L = 1.0\cdot 10^{-2} \cdot 10^{-2} \pm 3\cdot 10^{-3} \cdot 10^{-2}$. We observe that the logical failure rates using the morphed $[[10, 1, 2]]$ code are of similar order of magnitude as this state-of-the-art magic state preparation on the Steane code~\cite{postler2022demonstration, beverland2021cost}.

The logical failure rates shown in Fig.~\ref{fig:projected_performance} are averaged over different initial input states. However, the number of Pauli X- and Z-error configurations that cause a logical failure may vary, depending on the direction of switching and the given input state. Fig.~\ref{fig:projected_performance_variants}(a) shows the failure rates for the deterministic Hadamard- and T-gate for ideal logical input states $|\overline{0}\rangle$ and $|\overline{+}\rangle$. 
We observe that the failure rates for the logical input state $|\overline{+}\rangle$ are higher than for the input state $|\overline{0}\rangle$. Considering the Hadamard-gate, this can be explained by looking at the dangerous Pauli error positions on data qubits. As discussed in Sec.~\ref{sec:FT_CS}, Pauli Z-errors on data qubits can induce a logical failure when switching from the tetrahedral $[[15, 1, 3]]$ code to the $[[7, 1, 3]]$ Steane code. Analogously, Pauli X-errors on data qubits can cause a logical failure when switching from the Steane code to the tetrahedral code. For the initial state $|\overline{0}\rangle$, Pauli Z-errors do not affect the encoded state, since Z$|0\rangle = |0\rangle$. After switching to the Steane code, the state is changed to $|\overline{+}\rangle$ by applying the Hadamard-gate on the Steane code. The dangerous Pauli X-errors during the second switching step do not affect this encoded state. So, the Hadamard-gate on $|\overline{0}\rangle$ on the tetrahedral code is inherently less sensitive to these errors on data qubits than on the initial state $|\overline{+}\rangle$. Furthermore, $|\overline{0}\rangle$ is an eigenstate of the T-gate, so it is expected that logical failure rates are lower than for other states, in agreement with recent experimental observations~\cite{postler2022demonstration}. 

To estimate the impact of an imperfect initialization, we determine the logical failure rates for the same protocols including a noisy initialization as shown in Fig.~\ref{fig:projected_performance_variants}(b). The initializations are implemented using the circuits shown in Fig.~\ref{fig:init_tetra_logical_0_circuit}, \ref{fig:init_tetra_logical_plus_circuit}, \ref{fig:encoding_bulk}, \ref{fig:10_code_circuits} and \ref{fig:Steane_init}. Including a noisy initialization increases $p_L$ by a factor of approximately $ 1.1$. This indicates that the logical failure rate is dominated by the noisy switching procedure, rather than by the noisy initialization of logical input states. 

We identify the parameters $p_2$ as a limiting factor for the above code switching schemes. It is the largest value in our noise model and the number of two-qubit gates is much higher than for any other component. In order to estimate the sensitivity of the considered protocols to changes in $p_2$, we choose $p_1 = p_i = p_m = 0$ and only vary the two-qubit gate parameter $p_2$. Fig.~\ref{fig:p2_sensitivity} shows the averaged logical failure rates for the composite protocols as specified in Fig.~\ref{fig:protocols_overview}(c) for this noise model. The logical failure rate for the initial value of $p_2 = 3\cdot 10^{-3}$ for the Hadamard-gate is $p_L = 2.4 \cdot 10^{-2}$ and for the T-gate $p_L = 2.5 \cdot 10^{-2}$. This is close to the logical failure rates for the above noise model, where we included errors on single-qubit operations. This is in alignment with the expectation that faulty single-qubit operations have a minor impact on the logical failure rate and the probability for errors on two-qubit gates $p_2$ is the limiting parameter. Furthermore, we observe that an improvement of a factor of $5$ should suffice to reach the regime of the corresponding physical error rate for a FT non-deterministic magic state preparation, as shown in Fig.~\ref{fig:p2_sensitivity}(b) on the left. 

Overall, we report that the morphed $[[10, 1, 2]]$ code can reduce the logical failure rates by up to two orders of magnitude compared to the tetrahedral $[[15, 1, 3]]$ code, as for example for magic state preparation shown in Fig.~\ref{fig:projected_performance}. 
Furthermore, the FT deterministic logical gates are within reach of recently proposed implementations~\cite{postler2022demonstration, beverland2021cost, chamberland2019fault}. 

\section{Conclusions and Outlook}\label{sec:conclusion}

In this work, we have constructed fault-tolerant (FT) code switching protocols that enable the implementation of a FT universal gate set on the logical level. We provide specific instructions for the implementation of code switching protocols based on FT building blocks offering multiple methods as summarized in Fig.~\ref{fig:protocols_overview}. 
Composite code switching protocols can be constructed from this suite by composing building blocks in ways that are most suitable and optimized for given experimental setups. 
Based on current ion trap quantum processors, we estimate the performance of code switching protocols on near-term devices using a multi-parameter noise model and find that the logical failure rates reach values on par with state-of-the-art magic state injection~\cite{andersen2020repeated, postler2022demonstration}. 

A study comparing the overhead and thresholds for preparing magic states on the Steane code using magic state distillation (MSD) and deterministic code switching does not find an advantage in using code switching over MSD for this task~\cite{beverland2021cost}. 
We complement this study by considering morphed codes. 
As demonstrated in our work, stabilizer weights and, therefore, the circuit depth of a given protocol can be reduced by replacing the distance-three tetrahedral color code with the respective morphed code. This can be generalized by morphing tetrahedral color codes with higher distances~\cite{vasmer2022morphing} and might offer a more feasible alternative for magic state preparation on larger codes. 

Our numerical results provide a basis for the experimental realization of code switching protocols on existing or near-term quantum processors. Our protocols are readily implementable in architectures offering all-to-all qubit connectivity, such as e.g.~in ion trap quantum processors~\cite{kaushal2020shuttling, pino2021demonstration}. It would be interesting to adapt the protocols and also consider connectivity of other platforms, in particular nearest neighbor connectivity of superconducting qubits~\cite{bravyi2022future} or dynamically reconfigurable quantum registers of neutral atom processors~\cite{henriet2020quantum, evered2023high} or shuttling-based trapped-ion architectures~\cite{moses2023race, kaushal2020shuttling}. 
We emphasize that the proposed protocols and quantum circuit constructions are not limited to low-distance codes, but can be readily leveraged to scalable larger-distance two- and three-dimensional color codes. Furthermore, tailoring FT code switching to setups with biased noise~\cite{pal2022relaxation, huang2022tailoring, jain2023improved} offers the potential of reaching even lower logical failure rates for suitable experimental systems with such noise characteristics.

\section*{Acknowledgements}\label{sec:acknowledgements}

We would like to thank Josias Old for valuable discussions on FT encoding circuits. We gratefully acknowledge support by the EU Quantum Technology Flagship grant under Grant Agreement No.820495 (AQTION), the European Union’s Horizon Europe research and innovation programme under grant agreement No. 101114305 (“MILLENION-SGA1” EU Project), the U.S. Army Research Office through Grant No. W911NF-21-1-0007, the European Union’s Horizon Europe research and innovation program under Grant Agreement No. 101046968 (BRISQ), the ERC Starting Grant QNets through Grant No. 804247, by the Deutsche Forschungsgemeinschaft (DFG, German Research Foundation) under Germany’s Excellence Strategy “Cluster of Excellence Matter and Light for Quantum Computing (ML4Q) EXC 2004/1” 390534769. This research is also part of the Munich Quantum Valley (K-8), which is supported by the Bavarian state government with funds from the Hightech Agenda Bayern Plus.
This research is also supported by the Office of the Director of National Intelligence (ODNI), Intelligence Advanced Research Projects Activity (IARPA), via the U.S. Army Research Office through Grant No. W911NF-16-1-0070. The views and conclusions contained herein are those of the authors and should not be interpreted as necessarily representing the official policies or endorsements, either expressed or implied, of the ODNI, IARPA, or the U.S. Government. The U.S. Government is authorized to reproduce and distribute reprints for governmental purposes notwithstanding any copyright annotation thereon. Any opinions, findings, and conclusions or recommendations expressed in this material are those of the author(s) and do not necessarily reflect the view of the U.S. Army Research Office. The numerical simulations were performed with the aid of computing resources at Forschungszentrum J\"ulich.

\bibliography{references}

\clearpage
\appendix
\section{Morphing the tetrahedral $[[15, 1, 3]]$ code into the $[[10, 1, 2]]$ code}\label{app:morphing}

\begin{figure*}[b]
     \centering
     \begin{subfigure}[b]{0.9\textwidth}
         \centering
         \includegraphics[width=\textwidth]{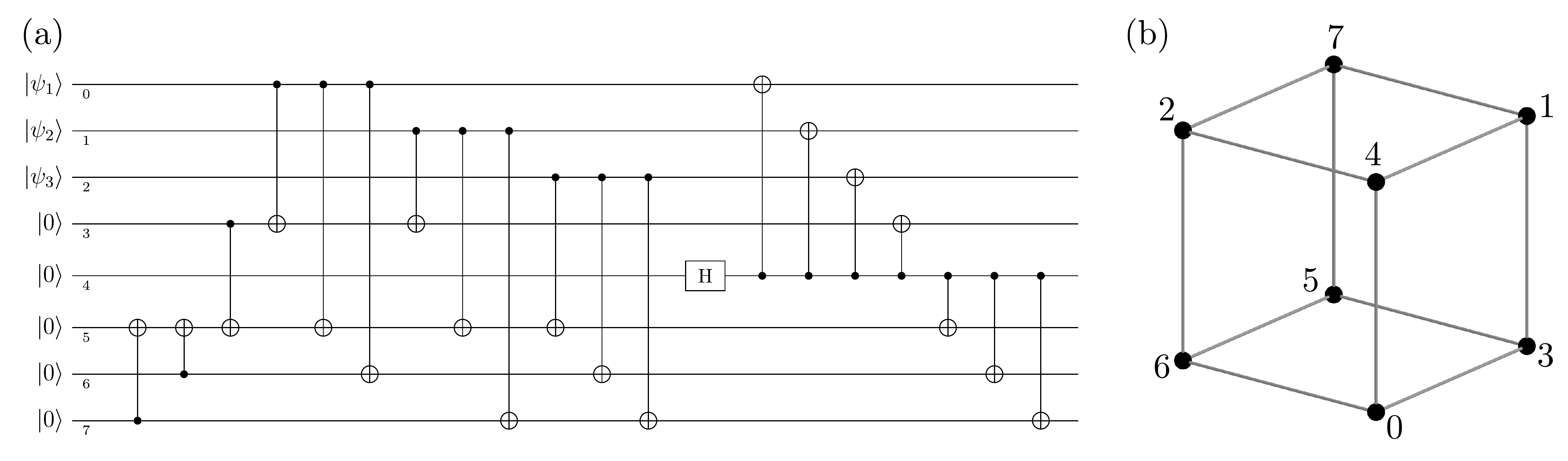}
     \end{subfigure}
    \caption{\justifying \textbf{Encoding of a logical state $|\overline{\psi_1}\, \overline{\psi_2}\, \overline{\psi_3}\rangle $ on the $[[8, 3, 2]]$ code.} We construct this circuit by initializing a $+1$-eigenstate of all stabilizers, as given in Eq.~\ref{eq:child_stabilizers} and coupling qubits $0, 1$ and $2$, which are in state $|\psi_1\rangle$, $|\psi_2\rangle$ and $|\psi_3\rangle$, to the qubits to the corresponding logical operator. }
    \label{fig:encoding_child_code}
\end{figure*}

\begin{figure*}[tb]
     \centering
     \begin{subfigure}[b]{0.7\textwidth}
         \centering
         \includegraphics[width=\textwidth]{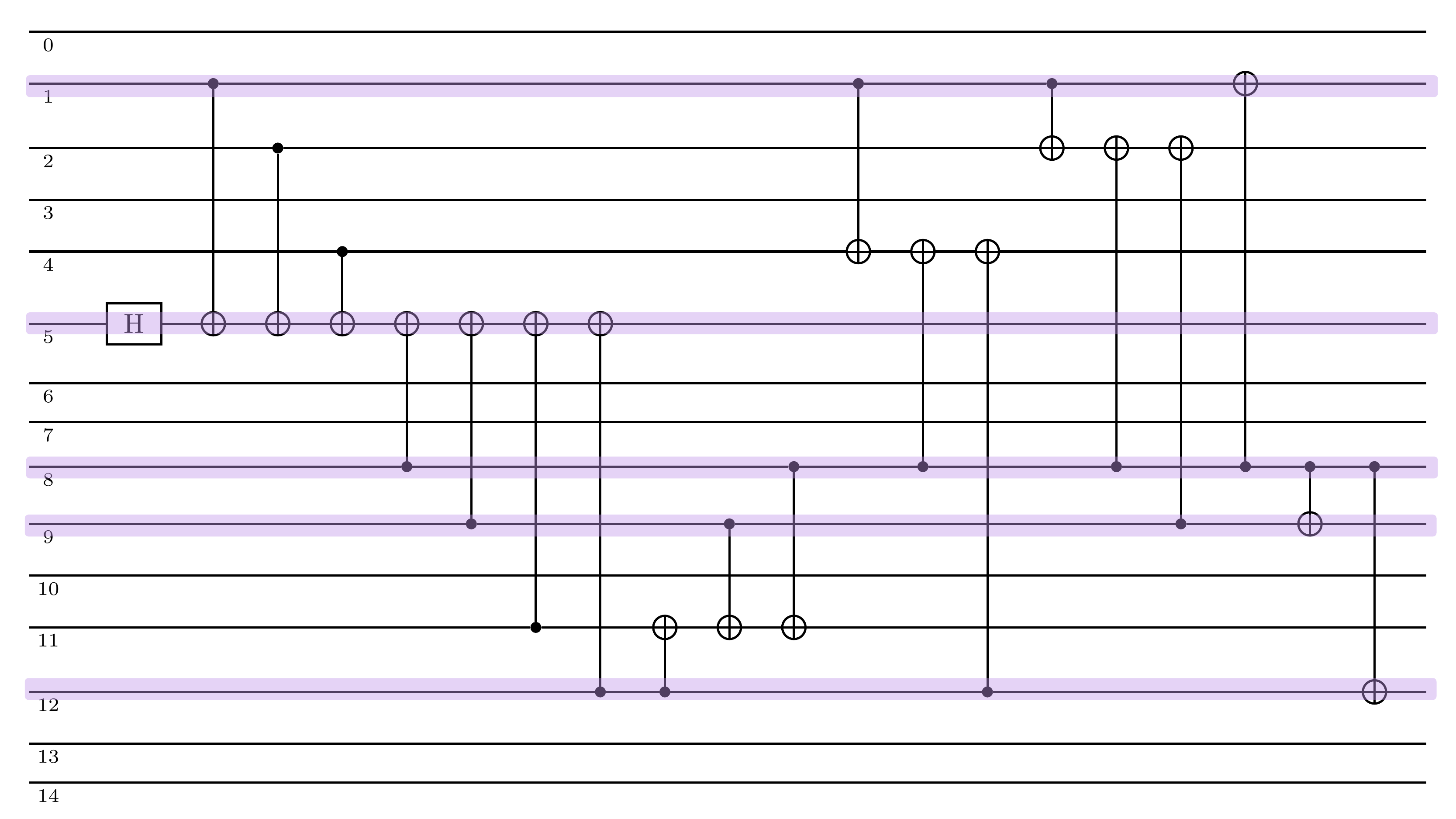}
     \end{subfigure}
    \caption{\justifying \textbf{Circuit for morphing the $[[15, 1, 3]]$ code into the $[[10, 1, 2]]$. } We find this circuit by inverting the encoding circuit of the child $[[8, 3, 2]]$ code and applying it to the yellow cell of the tetrahedral code for the indexing specified in Fig.~\ref{fig:Steane_and_tetra_code}(b). The marked qubits are effectively decoupled from the remaining $10$ qubits. }
    \label{fig:morphing_tetra_circuit}
\end{figure*}

The parent $[[15, 1, 3]]$ code defined on the tetrahedron contains a smaller sub-code on the yellow cell, which is a $[[8, 3, 2]]$ stabilizer code. The stabilizers of the $[[8, 3, 2]]$ code are~\cite{campbell2016smallest}
\begin{align}
	S_X &= X_0 X_1 X_2 X_3 X_4 X_5 X_6 X_7,  \\ 
	S_Z^1 &= Z_0 Z_1 Z_2 Z_3 Z_4 Z_5 Z_6 Z_7,  \nonumber\\
	S_Z^2 &= Z_0 Z_2 Z_4  Z_6 , \nonumber \\  S_Z^3 &= Z_1 Z_2 Z_4 Z_7,  \nonumber \\ S_Z^4 &= Z_0 Z_1 Z_3 Z_4. \nonumber \label{eq:child_stabilizers}
\end{align}
for the indexing given in Fig.~\ref{fig:encoding_child_code}(b). The logical operators are given by
\begin{align}
	\overline{X_1} &= X_0 X_3 X_5 X_6, \quad \overline{Z_1}= Z_0 Z_4  \\
	\overline{X_2} &= X_1 X_3 X_5 X_7, \nonumber \quad \overline{Z_2}= Z_1 Z_4 \\
	\overline{X_3} &= X_2 X_5 X_6 X_7, \quad \overline{Z_3}= Z_2 Z_4. \nonumber\label{eq:child_logicals}
\end{align}
A logical state $|\overline{\psi_1} \,\overline{\psi_2} \,\overline{\psi_3} \rangle $ can be encoded using the circuit shown in Fig.~\ref{fig:encoding_child_code}. 
We invert this encoding circuit by reversing the ordering of the applied operations and the direction of all CNOT-gates. We apply this inverted circuit to a codestate of the $[[15, 1, 3]]$ parent code, as shown in Fig.~\ref{fig:morphing_tetra_circuit}. In doing so, we effectively decouple $5$ qubits, meaning that they are not entangled with the remaining $10$ qubits, while the initial information is still encoded on these $10$ remaining qubits. It is important to note that the circuit shown in Fig.~\ref{fig:morphing_tetra_circuit} is not physically executed, since it is meant to generate a new code out of an existing one. For our protocols, we use the obtained morphed $[[10, 1, 2]]$ code directly. \\

\section{Encoding circuits}
\begin{figure*}[t]
     \centering
     \begin{subfigure}[b]{0.6\textwidth}
         \centering
         \includegraphics[width=\textwidth]{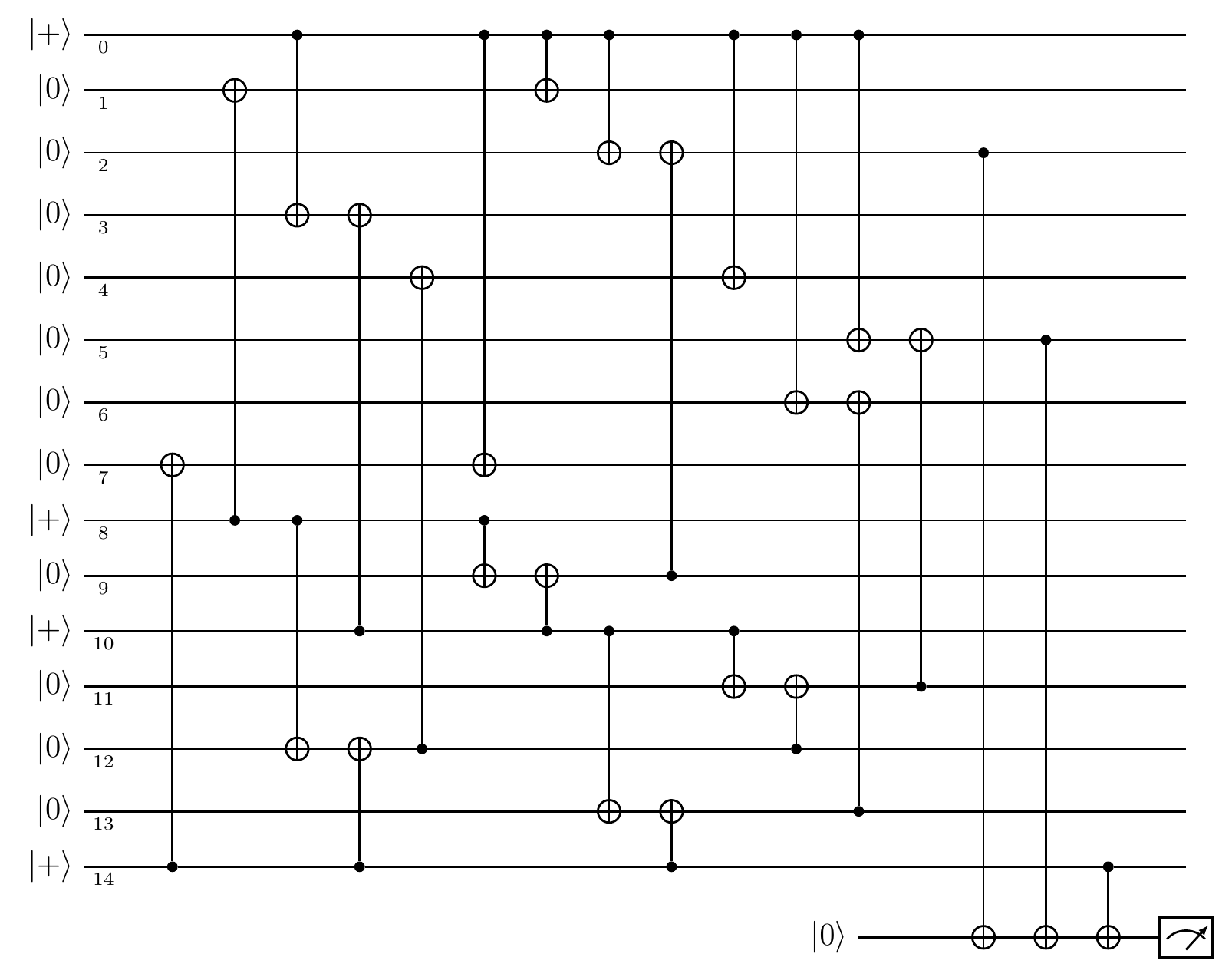}
     \end{subfigure}
    \caption{\justifying \textbf{Circuit for the initialization of $\ket{\overline{0}}$ on the $15$-qubit tetrahedral code $[[15, 1, 3]]$.} We obtain this circuit using the Latin rectangle method~\cite{steane2002fast}. A verification step is added in order to achieve fault tolerance. This verification corresponds to the measurement of a logical $\overline{Z} = Z_2 Z_5 Z_{14}$ and detects all incorrectable weight-four $X$-errors that result from a single propagated error. In total, $25$ CNOT-gates and $16$ qubits are required to fault-tolerantly initialize $\ket{\overline{0}}$.}
    \label{fig:init_tetra_logical_0_circuit}
\end{figure*}

\begin{figure*}[t]
     \centering
     \begin{subfigure}[b]{0.85\textwidth}
         \centering
         \includegraphics[width=\textwidth]{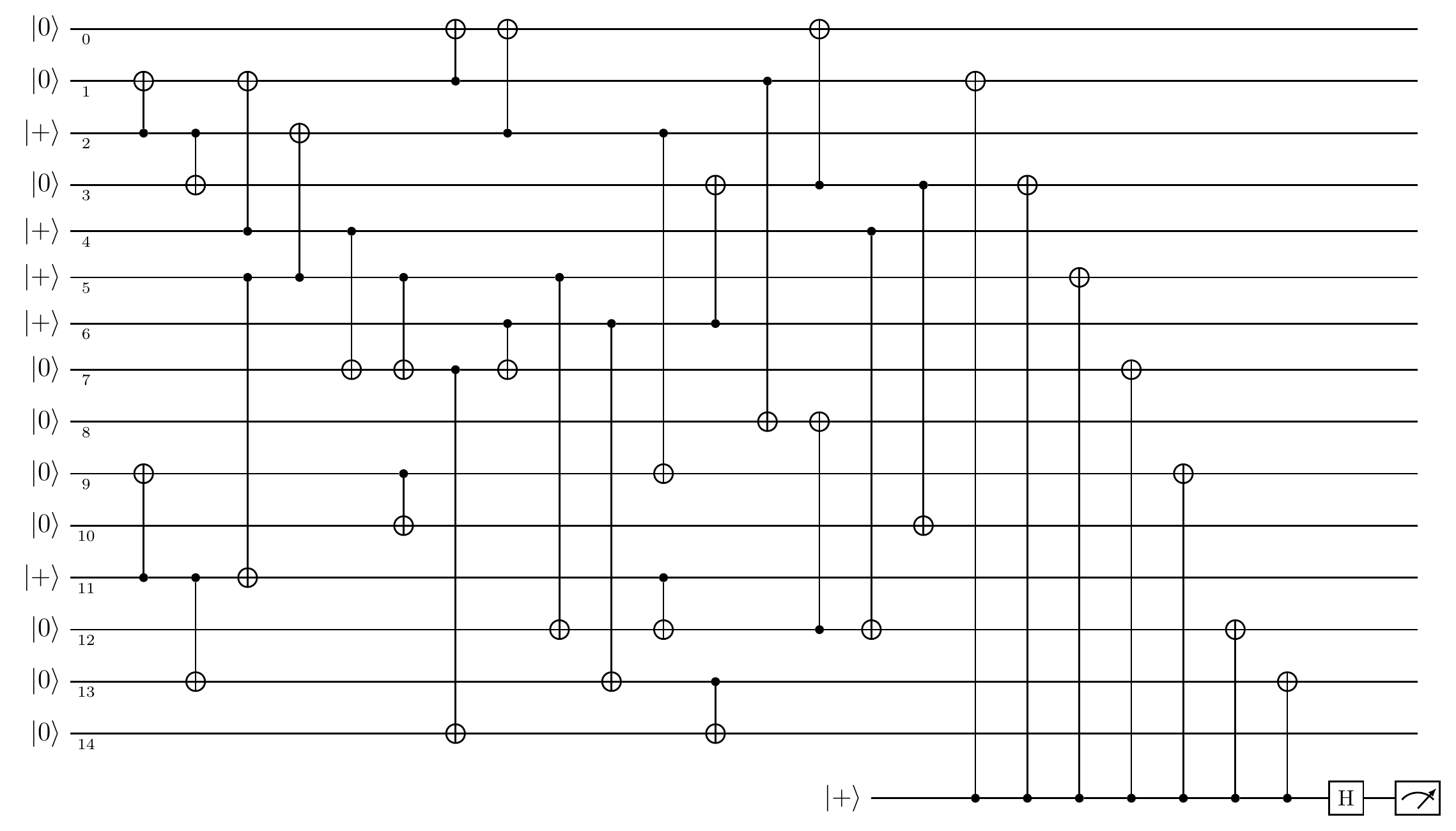}
     \end{subfigure}
    \caption{\justifying \textbf{Circuit for the initialization of $\ket{\overline{+}}$ on the $15$-qubit tetrahedral code $[[15, 1, 3]]$.} We construct this circuit using the Latin rectangle method~\cite{steane2002fast}. In this case, the verification step corresponds to the measurement of a logical $\overline{X} = X_1 X_3 X_5 X_7 X_9 X_{12} X_{13}$ and it detects all incorrectable weight-two $Z$-errors that result from a single propagated error. In total, $32$ CNOT-gates and $16$ qubits are required to fault-tolerantly initialize $\ket{\overline{+}}$. }
    \label{fig:init_tetra_logical_plus_circuit}
\end{figure*}

\begin{figure*}[t]
     \centering
	\includegraphics[width=\textwidth]{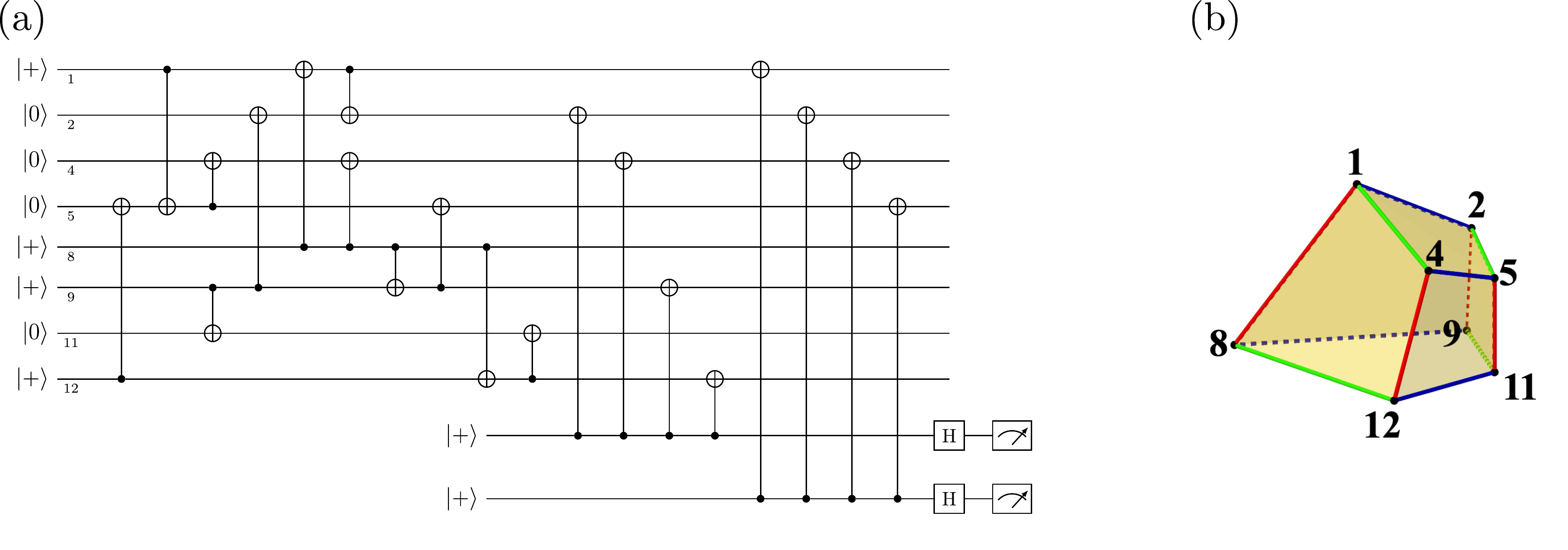}
    \caption{\justifying \textbf{Circuit for preparing the bulk for code switching.} (a) Encoding circuit of the initialization of the bulk (yellow cell) for switching from the Steane code to the tetrahedral $[[15, 1, 3]]$ code. Since the bulk has to be prepared in a specific state but does not encode any information, there is no logical operator that can be used in a verification step. Therefore, two stabilizers have to be measured in order to verify the resulting state. We choose the stabilizers $X_1 X_4 X_8 X_{12}$ and $X_1 X_2 X_4 X_5$ in order to detect weight-two Z-errors on the data qubits, which would result in a logical failure on the target $[[15, 1, 3]]$ code. (b) Indexing of the bulk qubits used for the encoding circuit in alignment with the previous sections. }
    \label{fig:encoding_bulk}
\end{figure*}

\begin{figure*}[t]
     \centering
	\includegraphics[width=\textwidth]{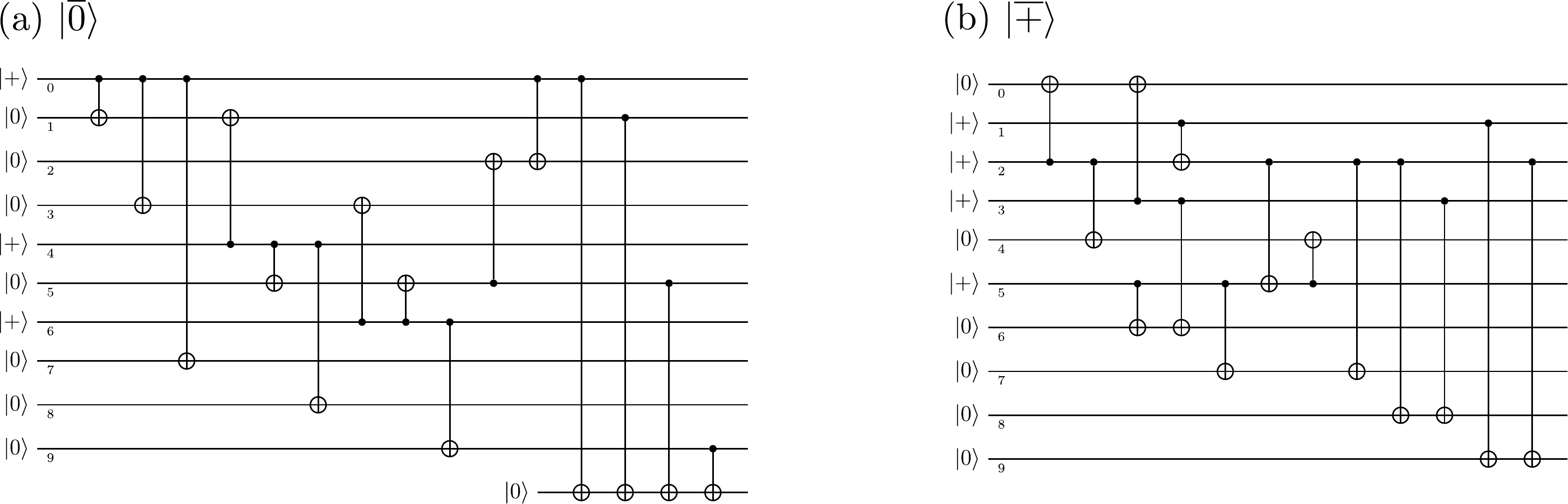}
    \caption{\justifying \textbf{Encoding circuit for the FT initialization of the morphed $[[10, 1, 2]]$ code. } Circuits for the FT initialization of (a) $\ket{\overline{0}}$ and (b) $\ket{\overline{+}}$ on the morphed $[[10, 1, 2]]$ code. For the FT initialization of $\ket{\overline{0}}$, the logical operator $\overline{Z} = Z_0 Z_1 Z_5 Z_9$ is measured to verify the resulting state. No verification step is required for the FT initialization of $\ket{\overline{+}}$, because every propagated Z-error is detectable afterwards on the $[[10, 1, 2]]$ code. }
    \label{fig:10_code_circuits}
\end{figure*}

\begin{figure*}[t]
    \centering
    \includegraphics[width=0.6\textwidth]{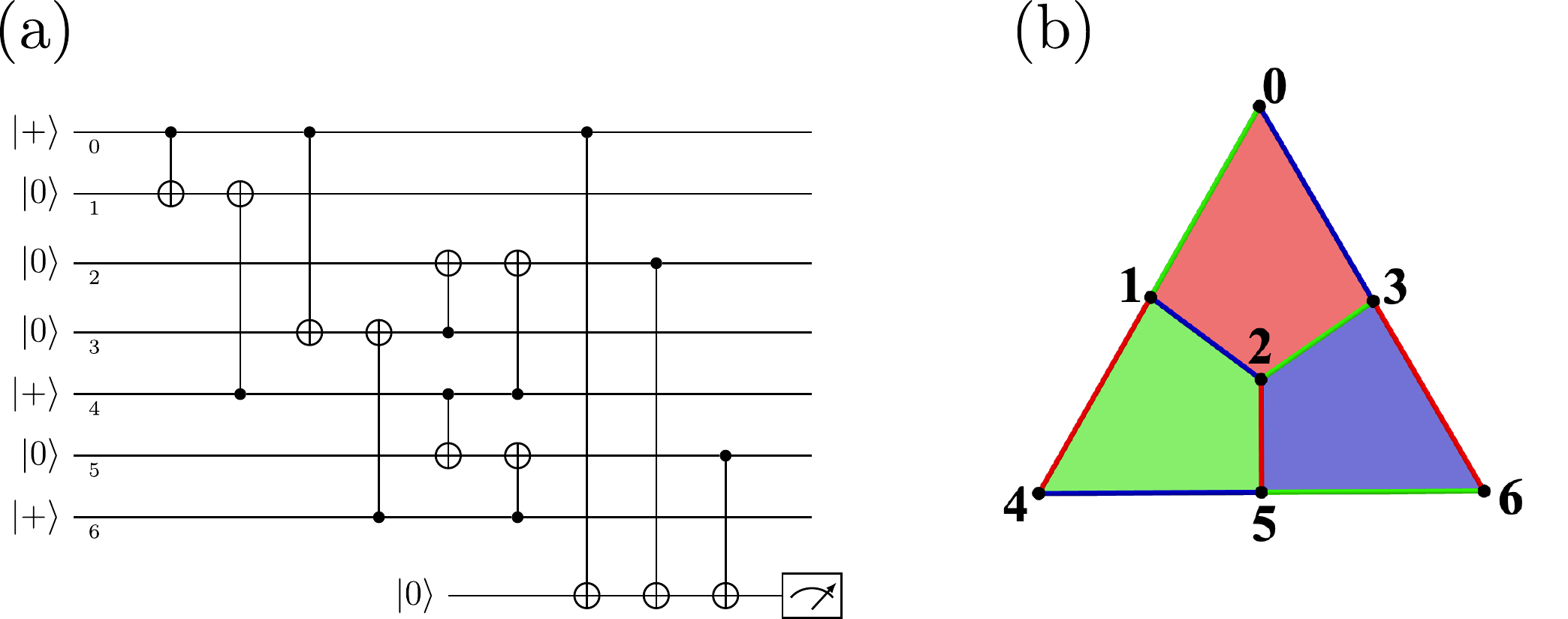}
    \caption{\justifying \textbf{Encoding circuit for the FT initialization of the $[[7, 1, 3]]$ Steane code~\cite{goto2016minimizing, postler2022demonstration}.} The logical operator $\overline{Z} = Z_0 Z_2 Z_5$ is measured in order to detect all single errors that propagate and would cause a logical failure. }
    \label{fig:Steane_init}
\end{figure*}

The following circuits shown in Figs.~\ref{fig:init_tetra_logical_0_circuit}--\ref{fig:Steane_init} can be used to implement logical states $\overline{\ket{0}}$ and $\overline{\ket{+}}$ on the tetrahedral code $[[15, 1, 3]]$ and the morphed $[[10, 1, 2]]$ code. We obtain the circuits using the Latin rectangle method~\cite{steane2002fast}. 

\end{document}